\documentclass{emulateapj}

\slugcomment{accepted for publication by the Astrophysical Journal}

\shortauthors{Heller, Shlosman and Athanassoula}
\shorttitle{Structure Formation in CDM Halos}

\begin{document}

\def\gtorder{\mathrel{\raise.3ex\hbox{$>$}\mkern-14mu
     \lower0.6ex\hbox{$\sim$}}}
\def\ltorder{\mathrel{\raise.3ex\hbox{$<$}\mkern-14mu
     \lower0.6ex\hbox{$\sim$}}}

\title{Structure Formation Inside Triaxial Dark Matter Halos:\\ 
Galactic Disks, Bulges and Bars}

\author{
Clayton H.\, Heller,\altaffilmark{1} Isaac\, Shlosman\altaffilmark{2} and
E.\, Athanassoula\altaffilmark{3}
}

\altaffiltext{1}{Department of Physics, Georgia Southern University,
    Statesboro, USA}
\altaffiltext{2}{Department of Physics \& Astronomy,
University of Kentucky, Lexington, USA}
\altaffiltext{3}{Observatorie de Marseille, Marseille, France}

\begin{abstract}
We investigate the formation and subsequent co-evolution of galactic disks immersed in 
assembling live dark matter (DM) halos. Disk and halo structural components have been 
evolved from the cosmological initial conditions and represent the collapse of an isolated
density perturbation. The baryonic component includes gas (which participates in star
formation [SF]) and stars. The feedback from the stellar energy release onto the ISM 
has been implemented, and so is the mass, momentum and energy 
balance between the stellar and gaseous components. We find that
(1) The growing {\it triaxial} halo figure tumbling is insignificant and the angular momentum
($J$) is channeled into the internal circulation, while the baryonic collapse is stopped by
the centrifugal barrier; (2) Density response of the (disk) baryonic component is out of 
phase with the DM, thus diluting the inner halo flatness and washing out its prolateness
over the time period of the disk growth; (3) The total $J$ is neathly conserved, even in 
models accounting for feedback from stellar evolution; i.e., the baryons lose $\sim 
25\%-30\%$ and the DM acquires $\sim 2.5\%-3\%$ of their original $J$; baryons
and DM within the disk radius have the same $J$ after the first $\sim 3$~Gyr; 
(4) The specific
$J$ for the DM is nearly constant, while that for baryons is decreasing with time; 
(5) Early stage of disk formation resembles the cat's cradle --- a small amorphous disk 
fueled via radial string patterns; followed by growing oval disk whose shape varies with
its orientation to the halo major axis, and so is the strength of a pair of 
grand-design arms developed in the disk; (6) The initially puffed up gas component in 
the disk thins when the SF rate drops below $\sim 5~{\rm M_\odot~yr^{-1}}$; 
(7) About 40\%-60\% of the baryons remain outside the SF region of the disk
or in the halo in the form of a hot gas by the end of the simulations;
(8) Obtained rotation curves appear to be flat and account for the observed disk/halo
contributions (the halo virial mass is $\sim 7\times 10^{11}~{\rm M_\odot}$
at $z=0$);
(9) A range of bulge-dominated to bulgeless disks was obtained, depending on the energy 
thermalization parameter, $\epsilon_{\rm SF}$, which characterizes the feedback from the 
stellar evolution --- smaller  $\epsilon_{\rm SF}$ leads to a larger and earlier bulge; 
Lower density threshold for SF leads to a smaller, thicker disk; gravitational
softening in the gas has a substantial effect on various aspects of galaxy evolution
and mimics a number of intrinsic processes within the interstellar medium;
(10) The models are characterized 
by an extensive bar-forming activity within the central few kpc whose properties vary with 
the bar-halo orientation; 
(11) Nuclear bars, dynamically coupled and decoupled form in 
response to the gas inflow along the primary bars, as shown by Heller, Shlosman \&
Athanassoula. 
\end{abstract}
 
\keywords{galaxies: evolution -- galaxies: formation -- galaxies: halos -- galaxies: 
kinematics and dynamics -- galaxies: structure -- cosmology: dark matter}

\section{Introduction} 

Hierarchical Cold Dark Matter (CDM) cosmology has been largely
successful in establishing the paradigm of structure formation and evolution
on spatial scales of $\sim 100$~kpc, corresponding to individual halos, and
beyond (e.g., Cole et al. 2000; Steinmetz \& Navarro 2002; Spergel et al.
2003). However, on smaller scales, a number of fundamental issues
related to the origin of luminous galactic components and their structural
parameters remain unresolved because of our limited understanding of 
dynamical processes, such as mass and angular momentum redistribution, star 
formation (SF) and the energy and momentum feedback from the stellar evolution 
onto the interstellar medium (ISM). Specifically, no concensus
exists at present on formation of galactic disks, bulges, bars and supermassive
black holes (SBHs, e.g., White \& Rees 1978; Fall \& Efstathiou 1980; White 1996; 
Ellis 2000). Because these structures are made from the baryonic
matter, either gas or stars or their mixture, a certain degree of dissipation 
must be involved. Detection of quasars at redshifts $z\gtorder
6$ (e.g., Fan et al. 2001; Mathur, Wilkes \& Gosh 2002; Bertoldi et al. 2003), 
when the age 
of the Universe was less than a gigayear, rises intriguing
questions about the origin of their host galaxies within the framework of the 
hierarchical clustering, and is important for the fate of the SBH ---
bulge correlation, i.e., $M_\bullet-M_{\rm bulge}$, and the masses 
of the quasar host galaxies, at these $z$ (e.g., Trenti \& Stiavelli 2007).
In this work we investigate how various dissipative processes, including the SF
 and the feedback from stellar evolution onto the ISM, affect the galactic 
disk and bulge formation and evolution. For
this purpose we follow a monolithic collapse of an initial perturbation in
the DM and baryons within the Hubble flow.  

When exactly did the disks and other structural elements of disk galaxies, like
bulges, bars and central SBHs, assemble? How much evolution 
has occured since? Both orbital and ground-based telescopes have began to 
tighten the constraints on these issues, but essential details still missing and
controversies are abound. First, within the hierarchical merging scenario,
the so-called `angular momentum catastrophe' --- the angular momentum
($J$) distribution of the collapsing baryonic matter does not match that of the
observed disks in nearby Universe (e.g., Sommer-Larsen, Gotz \& Portinari 2002; 
Maller \& Dekel 2002; van den Bosch 2002b; Burkert \& D'Onghia 2004).
This distribution is characterized by low and high $J$ tails which result both
in the over-extended disks and very massive bulges. In addition, gravitational
torques between the collapsing baryons and the DM halos lead to 
disks with radial scalelengths substantially smaller than those observed.
Under the assumption of $J$ conservation during the collapse, the specific
$J$ generated by the random work in the expanding Universe
is too small to explain the formation of bulgeless disks in the late type 
galaxies (van den Bosch, Burkert \& Swaters 2001; van den Bosch 2002a). 

The second issue is that of the `core catastrophe' (e.g., Moore et al. 1999)
and relates to the DM central density distribution. The numerical simulations of DM 
halo formation have shown that these attain a universal density profile which can 
be nicely fitted by a two-parameter family (Navarro et al. 1996, hereafter NFW). 
However, observations of disk galaxies
and galaxy clusters indicate that the DM halos have flat density cores rather
than NFW cusps (e.g., Flores \& Primack 1994; Moore 1994; Burkert 1995; Kravtsov
et al. 1998; Salucci \& Burkert 2000; Boriello \& Salucci 2001; de Blok \& Bosma 2002; 
Sand et al. 2004). The standard picture of slow adiabatic collapse of a smooth 
baryonic component in the centers of such halos can only lead to a larger central 
concentration, making matters worse. El-Zant, Shlosman \& Hoffman
(2001) and El-Zant et al. (2004) have shown that a flattening of the central
density cusps can be a natural outcome of energy deposition by the accreted clumpy 
baryons via a dynamical friction both in galaxies and in clusters of galaxies.
Alternatively, Weinberg \& Katz (2002) and Holley-Bockelmann, Weinberg, Katz (2005) 
have claimed that the cusps can be leveled through the action of the stellar bars, 
but this was strongly disputed by Sellwood (2003) and McMillan \& Dehnen (2005). 

Third, Abraham et al. (1999) and van den Bergh et al. (2002) have claimed that
the fraction of stellar bars
sharply decreases above the redshifts of $z\sim 0.5$, while the analysis of
GEMS survey (Rix et al. 2004) of 1600 galaxies has shown that both the
fraction of bars and their size and axial ratio distributions remain largely
unchanged up to $z\sim 1$ at least (Jogee et al. 2004). This latter evidence
is supported by Elmegreen et al. (2004) and Sheth et al. (2003) based on much
smaller samples of 186 and 4 galaxies respectively. 

Finally, the DM halos resulting from cosmological simulations appear 
to be strongly triaxial, both in their prolateness and flatness,\footnote{We 
define the density prolateness as $\epsilon_\rho=1-b/a$, where $b/a$ is the
intermediate-to-major axis ratio and the flatness as $f_\rho=1-c/a$ 
with $c/a$ --- the minor-to-major axis ratio in the DM halo}
$\epsilon_\rho$ and $f_\rho\sim 0.5$ (e.g., Allgood et al. 2006). On the other 
hand, halo shapes inferred from observations in the local Universe are
axisymmetric, although individual objects may exhibit some mild degree 
of prolateness (e.g., Rix \& Zaritsky 1995; Merrifield 2002). Theoretical arguments 
supplemented by numerical simulations have shown that bars are incompatible 
with prolate halos (El-Zant \& Shlosman 2002; Berentzen, Shlosman \& Jogee 2006).
When taken in tandem with the plethora of stellar bars observed locally (e.g., 
Marinova \& Jogee 2007; Menendez-Delmestre et al. 2007; Knapen, Shlosman \& 
Peletier 2000) and at the intermediate redshifts of $\sim
1$, corresponding to the lookback time of 8~Gyr (Jogee et al. 2004), these
results indicate strongly that the halo shapes have evolved only
mildly during this time period. Furthermore, Dubinski (1994) and Kazantzidis
et al. (2004) have shown that the halos lose some of their triaxiality
when baryons are added. In the next step, Berentzen \& Shlosman (2006) have 
simulated the disk growth in
an assembling DM halo within the $4h^{-1}$~Mpc comoving box. The initial
conditions have been produced by means of the constrained realizations algorithm
of Hoffman \& Ribak (1991). The halo prolateness was found to be washed out
within the disk radius, and the halo flatness to decrease substantially.

The strongly nonlinear process of galaxy formation coupled with a long list of 
dissipative processes provide a strong insentive for a numerical
approach. A number of numerical methods to follow up the disk formation have
been attempted. In the models of Immeli et al. (2004), when the gas in the
proto-disk cools efficiently, the disk fragments and forms a number of massive
clumps of stars and gas. These clumps spiral in to the galactic center and
merge there to produce a strong starburst, as also shown by Shlosman \&
Noguchi (1993), leaving a central object which
resembles a classical bulge. The disk growth was simulated in a fixed 
spherically-symmetric external 
potential of the DM halo by adding the gas directly to the
disk. Alternatively, Samland \& Gerhard (2003) have modeled a disk and 
the associated 
DM halo growth by adopting the mass inflow rate from {\it separate} cosmological 
simulations, while preserving the spherical symmetry of the halo, using a
chemo-dynamical code. A number of issues have been investigated, among them the
stellar population of the bulge. A possibility of forming bars from baryonic
spheroids immersed in triaxial DM halos was considered by Gadotti \& de Souza 
(2003; see also El-Zant et al. 2003).

Sommer-Larsen, G\"otz \& Portinari (2003) analyzed the formation history
of two galaxies with circular velocities comparable to the Milky Way (MW) and found gas 
accretion rates, and hence X-ray halo luminosities, at $z\sim 1$, 6--7 times larger than
at $z=0$ for these objects. More generally, it is found that the gas infall
rates onto these disks are nearly exponentially declining with time, both for
the total disk and for the ``solar cylinder.'' This result supports the exponentially
declining gas infall approximation often used in the chemical evolution models.

Okamoto et al. (2005) has investigated the effect of the SF feedback
on the morphological evolution of galaxies by assuming that the SF can proceed
in two modes --- either in the high density or shock-triggered fashion. The former 
mode leads to elliptical galaxies while the latter to
disk ones. The authors made a number of assumptions that require a further
validation, e.g., the SPH kernel was fixed at the constant mass and not on the
constant number of particles.

The simulations of Governato et al. (2004, 2007) form rotationally-supported disks
with realistic exponential scalelengths and fit both the $I$ band and baryonic
Tully-Fisher relations (see also Stinson et al. 2006; Kaufmann et al. 2007). An 
extended stellar disk forms inside the MW-sized
halo immediately after the first merger. The combination of the UV background and 
supernova (SN) feedback drastically reduces the number of visible satellites 
orbiting within this halo, bringing it in a fair agreement with observations. 
Formation of stellar populations in the MW-type galaxy were studied by Scannapieco 
et al. (2006) using a combination of high-resolution $N$-body simulations with
semi-analytical methods of metal enrichment. Substantial constraints on the 
formation of Population~III stars have been obtained. Brook et al. (2007b) have
followed up with chemodynamical simulations, including the SF, 
concluding that surveys of low-metallicity stars in the Galactic halo can 
directly constrain the properties of Population~III stars. Furthermore, the 
disk evolution was simulated by Brook et al. (2007a), examining the possibility 
that high $\alpha$ elements-to-iron abundance ratios detected in the thick-disk 
component of the Milky Way can be explained by SF during the gas-rich mergers.
 
All the above studies have brought a very useful insight into the formation of disk
galaxies and have contributed to a better understanding of
specific aspects of the problem. Yet, as already mentioned, the problem
is very complex and much more remains to be investigated, in particular
regarding the dynamics of forming disks, the interaction between the
various morphological components and subcomponents, as well as their origin and
properties. The present paper aims to advance of some of these open questions. 
Our emphasis is more on
the dynamical aspects and includes discussions of the formation
of bars, both inner and outer, made in such {\it ab initio} simulations.
Our algorithm includes the SF, the feedback
from the stellar evolution, etc. We also performed a
relatively large number of simulations, so as to gain insight on how
the SF and energy feedback affect the structural
properties of the galaxy components and subcomponents.
We use cosmological initial conditions and follow the Hubble expansion
and the subsequent collapse of a single perturbation in the gas and the DM
in order to analyze the role of different processes in the formation of disk
and bulge components. We aim at understanding the role of the external
cosmological and internal dissipative factors.

Section~2 describes the details of our 
method of modeling the SF, our approach in
determinig the shapes of spheroidal and disk components, and the initial
conditions used for numerical simulations. Sections~3 and 4 present
results of our simulations and model comparison. Discussion follows in the
last section.

\section{Numerical Methods}

\subsection{Modeling the Gas, Stars and DM}

The numerical simulations have been performed with an updated version of the
FTM-4.4 hybrid code which evolves the collisionless ($N$-body) and gaseous
(Smooth Particle Hydrodynamics, SPH) components (Heller \& Shlosman 1994;
Heller 1995), with $N\sim 5\times 10^5$ and $N_{\rm SPH}\sim 5\times 10^4$.
In addition, a number of models have been evolved using the parallel
version, FTM-4.5, of the code.
The gravitational forces were computed using the public version of the 
routine {\tt falcON} (Dehnen
2002), which is about ten times faster than optimally coded Barnes \& Hut
(1986) tree code. The tolerance parameter $\theta$ is fixed at 0.55.

The energy equation describing the internal state of the SPH
particles have been used to evolve a multiphase ISM by calculating its heating 
and cooling rates and iterating for the temperature. Star formation was used to 
convert SPH particles into collisionless
particles which exert an energy feedback onto the surrounding gas. The 
gravitational softening for the collisionless DM and stellar particles
is $\epsilon_{\rm *,grav} = 150$~pc, except for two models specified in
Table~1. For the SPH particles we use two options 
to soften the gravity --- constant softening of $\epsilon_{\rm g,grav} = 
150$~pc or 250~pc, and a dynamic softening (Heller \& Shlosman 1994) with the 
minimum of $\epsilon_{\rm g,min,grav} = 250$~pc.

A large number of tests has been performed to check the sensitivity of the
results to the algorithm and its parameters. A typical run 
with the DM only conserved the energy to within 1\% and $J$ to within 0.1\%.

\subsection{Star Formation}
 
We anticipate that the SF processes and the associated 
feedback from stellar evolution coupled with an overall dissipative and 
collisionless dynamics within the forming galaxy determine its evolutionary
path. Strictly speaking, because the smallest spatial scales resolved here
are substantially larger than scales which are relevant for an `individual' 
SF event, the underlying collapse should be treated witin the context of
the two-fluid Jeans instability (Zel'dovich \& Novikov 1975) in a sheared 
medium. We adopt the modified prescription for the SF from 
Heller \& Shlosman (1994) that required the gas to be Jeans unstable and 
subject to the feedback from OB stellar winds and supernovae (SN), and
introduced a number of 
statistical elements. Our main assumption is that the SF takes place in 
regions which are contracting, i.e., 
\begin{equation}
 {\bf \nabla }\!\cdot\!{\bf v} < 0,
\end{equation}
and are Jeans unstable, i.e.,  
\begin{equation}
  \tau_{\rm coll} < \tau_{\rm sound},
\end{equation}
which assures that the collapse would continue unhindered down to the 
smallest resolved spatial scales. Under this condition, the pressure 
gradients (which are established over the sound crossing time 
$\tau_{\rm sound}$) will not be able to build up during the collapse time
$\tau_{\rm coll}$. Moreover, if the gas cooling time satisfies 
$\tau_{\rm cool}\ll \tau_{\rm coll}$, this will ease the condition given by
eq.~(2), which, for simplicity, can be replaced by 
$\rho_{\rm gas} > \rho_{\rm crit} \equiv 7\!\times\!10^{-26}\,{\rm g~cm^{-3}}$
for temperatures in excess of $10^4$~K (e.g., Navarro \& White 1993). This
happens because the cooling function $\Lambda(T)/T$ has a minimum at $T\sim
1.6\times 10^6$~K.   

For larger densities than $\rho_{\rm crit}$, the short cooling time 
$\tau_{\rm cool}$ will be guaranteed if the {\rm neutral} (cold) gas will be 
at least moderately self-gravitating in the background density $\rho_{\rm back}$,  
\begin{equation}
  \rho_{\rm ngas}  > \alpha_{\rm crit}\rho_{\rm
      back}. 
\end{equation}
Here $\rho_{\rm ngas}$ corresponds for the neutral gas (i.e., H\,I and He\,I),
and $\rho_{\rm back}=\rho_{\rm hgas}+\rho_{\rm star}+\rho_{\rm DM}$ combines 
the mass of the {\rm ionized} (hot) gas, stars and DM, with 
$\alpha_{\rm crit}\ltorder 1$ being the fudge factor. For the standard model,
$\alpha_{\rm crit}= 0.5$ (see Table~1).
 
Gas is converted into stars at a rate 
\begin{equation}
\dot{\rho}_{\rm ngas} = -\frac{\rho_{\rm ngas}}{\tau_{\rm coll}}, 
\end{equation}
where the SF (collapse) timescale is taken as 
$\tau_{\rm coll}=\alpha_{\rm ff}\tau_{\rm dyn}$, with $\tau_{\rm dyn} =
\sqrt{3\pi/16\rho_{\rm ngas}}$. In reality, the collapse time for molecular
clouds is extended beyond the free-fall time, e.g., because of an additional 
support from the MHD turbulence. We have used $\alpha_{\rm ff}=10$ for the 
standard model.

Next, we introduce the probability that a gas particle of mass $m_{\rm g}$ 
produces a stellar particle of mass $m_{\rm s}$ during a given timestep of length 
$\Delta t$ of
\begin{equation}
p_{\rm SF} = \eta \left[ 1 - \exp\left( -\frac{f_{\rm ngas} \Delta t}
              {\tau_{\rm coll}} \right) \right],
\end{equation}
where
$f_{\rm ngas} = \rho_{\rm ngas}/\rho_{\rm gas}$ and $\eta = m_{\rm SF}/m_{\rm s}$.
Here $m_{\rm SF}= m_{\rm g} \left( 1 - R_{\rm SF} \right)$ represents the
stellar mass which is produced in the SF event.  However,
a stellar particle with a mass of only $m_{\rm s}=0.25 m_{\rm g}$ is created as $R_{\rm
SF}=0.4$ of the mass is instantaneously recycled back to the parent gas particle, 
along with an increment in its metallicity by $\Delta z_{\rm g} = Z_{\rm SF}m_{\rm
SF}$, where $Z_{\rm SF}=0.00793$ is the metals yield (Leitherer, Robert \& Drissen 1992). 
The quantity $\eta$ is not 
allowed to be less than unity, in which case we set $\eta=1$ and $m_{\rm s} = 
m_{\rm g}$, and the gas particle is removed from the simulation. 
Each gas particle is capable of creating four generations of stars with different 
metallicities and $m_{\rm g}$ decreases accordingly after each SF event.  
The mass of the stellar particle represents a cluster of stars. The fraction and timing
of massive stars that produce the OB line-driven winds and SN are calculated 
assuming the Salpeter IMF.

\subsection{Feedback from Stellar Evolution}

The thermal balance in the interstellar gas is calculated using the energy
equation. Sources of heating and cooling include adiabatic, viscous and radiative
processes. The balance of the specific internal energy along with the gas 
ionization fractions (H, H+, He, He+, He++, e) and the mean molecular weight are 
computed as a function of density and temperature for an assumed optically thin 
primordial composition gas (Katz et al. 1996).
This allows for the computation of the neutral gas density $\rho_{\rm ngas}$
required for the SF criteria of eqs.~(1--3).

To model the energy feedback from the stellar evolution onto the gas,
the newly formed stellar particles inject energy from SN and OB stellar 
winds into the $N_{\rm SF}=16$ surrounding gas particles. The total rate of energy 
deposition by stars, $E_{\rm SF} = E_{\rm wnd} + E_{\rm SN}$, is given by,
\begin{equation}
 E_{\rm wnd} = \epsilon_{\rm SF} P_{\rm wnd} m_{\rm SF} m_{\rm s}^{-1} \tilde{z}_{\rm s} ,   
\end{equation}
{\rm for} $t_{\rm s} < 6\!\times\!10^6\,{\rm yrs}$, and
\begin{equation}
 E_{\rm SN} = \epsilon_{\rm SF} P_{\rm SN} m_{\rm SF} 
\end{equation}
{\rm for} $3\!\times\!10^6\,{\rm yrs} <t_{\rm s} < 3\!\times\!10^7\,{\rm yrs}$.
Here $\epsilon_{\rm SF}$ is an energy thermalization parameter, 
$t_{\rm s}$ is the age of a stellar particle and 
$P_{\rm wnd} = P_{\rm SN} = 2.75\!\times\!10^{41}\,{\rm
erg~M_\sun^{-1}~yr^{-1}}$ are the energy deposition rates per solar mass
by massive stars and SN, respectively. The metallicity of a stellar
particle is given by $\tilde{z}_{\rm s}$ in units of the solar metallicity 
(Maeder 1992, 1993).
The timestep of such `active' stellar 
particles (and of all the gas particles) is restricted to being smaller than 
$\Delta t_{\rm fb}\equiv 6\!\times\!10^5$\,yrs in order to properly resolve the 
feedback timescale of $\tau_{\rm fb} = 3\!\times\!10^7$\,yrs.

The $N_{\rm SF}$ gas particles in the vicinity of the `active' stellar particles
are treated in the following way. Their radiative cooling is temporarily disabled 
when receiving the energy from a stellar particle and when the condition in 
eq.~(2) is still fullfilled.  The energy
$E_{\rm SF}$ is thermalized and deposited in the gas in 
the form of a thermal energy, then converted to kinetic energy through the equations 
of motion. This method is preferable over injecting a fraction of the stellar energy 
directly in the form of a kinetic energy --- this approach is ambiguous and the results
are sensitive to the value of a timestep and to the particle number on the receiving end. 

For a comparison, we have also run some models with an isothermal equation of state
(EOS), implementing the feedback by directly over-pressuring the gas through an 
increase of the local sound speed over the isothermal value.  Specifically the sound 
speed, $c_{\rm g}$, is increased by,
\begin{equation}
\Delta c_{\rm g} = E_{\rm SF} t_{\rm fb} \left(\gamma - 1\right)/2c_{\rm g}.
\end{equation}

\subsection{Halo Shape Determination}

To determine the intrinsic shape of the DM and stellar particle distributions, 
we remove any residual net velocity from the system, reject any unbound particles 
and iterate this procedure to locate the density center (e.g., Aguilar \& Merritt 1990), 
\begin{equation}
{\bf r}_{\rm d} = \frac{\sum \rho_i {\bf r}_i}{\sum \rho_i},
\end{equation}
where $\rho_i$ is the local mass density at the position of each
particle.  

In the reference frame defined by the so determined velocity centroid
and density center, we compute the eigenvalues of the moment of
inertia tensor for the mass within a specified radius.
From these we may determine the axes $a>b>c$ of a uniform
spheroid with the same eigenvalues.  The ratios of these axes may
then be used to characterize the shape of the system, as defined by
\begin{equation}
\frac{b}{a} = \sqrt{\frac{e_1-e_2+e_3}{\Delta}},
\end{equation}
\begin{equation}
\frac{c}{a} = \sqrt{\frac{e_1+e_2-e_3}{\Delta}},
\end{equation}
where $\Delta=e_2-e_1+e_3$ for the eigenvalues $e_3>e_2>e_1$.

\subsection{Initial Conditions}

The initial conditions are those of a spherically-symmetric density
enhancement in the Einstein-de~Sitter Universe. We use an open CDM (OCDM) 
model with $\Omega_0=0.3$, $h=0.7$, where $\Omega_0$ is the current cosmological density
parameter and $h$ is the Hubble constant normalized by
100~${\rm km~s^{-1}~Mpc^{-1}}$. Models with $\Omega_0=0.3$ and $\Lambda=0.7$ 
in $\Lambda$CDM Universe have been run as well, but will be discussed elsewhere.
Because the collapse time of our DM halos happens at $z\sim 2$ (Table~1), the
differences between the OCDM and $\lambda$CDM models are insignificant. We have
also rerun our standard model N3 with the WMAP3 baryon fraction of 17\%  (model
N41) and (separately) with 5 times larger number of the SPH and collisionless particles 
(model N42) --- the differences with the original model appear to be only quantitative.

\begin{deluxetable*}{lcccccccl}
\tablecaption{\bf MODELS}
\tablehead{
Model & $z_{\rm c}$ & $\epsilon_{\rm SF}$ & $\alpha_{\rm ff}$ & $\alpha_{\rm crit}$ & 
$\epsilon_{\rm *,grav}$ & $\epsilon_{\rm g,grav}$ & EOS & Notes}
\startdata

{\bf N\,3} & 2 & 0.3 & 10 & 0.5 & 0.015 & 0.015 &  & standard model: DM halo$+$baryons\\
{\bf QN\,3} & 2 & -- & -- & -- & 0.015 & -- &  & as N3 but DM only\\
\hline
 
{\bf N\,4} & 2 & 0.3 & 10 & 0.5 & 0.015 & 0.015 &  & enforced axisymmetric halo\\
\hline

{\bf N\,5} & 2 & 0.3 & 10 & 0.5 & 0.015 & 0.015 & isothermal &  \\
\hline
  
{\bf N\,6} & 2 & 0.05 & 10 & 0.25 & 0.015 & 0.015 & & \\
{\bf N\,8} & 2 & 0.1  & 10 & 0.25 & 0.015 & 0.015 & & \\ 
{\bf N\,14} & 2 & 0.3 & 1 & 0.5 &  0.015 & 0.015 & & \\
{\bf N\,15} & 2 & 0.1 & 1 & 0.5 & 0.015 & 0.015 & & \\
{\bf N\,16}& 2 & 0.1  & 1  & 0.25 & 0.015 & 0.015 & & \\
{\bf N\,17} & 2 & 0.1 & 10 & 0.5 & 0.015 & 0.015 & & \\
{\bf N\,18} & 2 & 0.3 & 10 & 0.5 & 0.025 & 0.025 & & \\
{\bf N\,19} & 2 & 0.1 & 1 & 0.5 & 0.025 & 0.025 & & \\
\hline
 
{\bf N\,20} & 2 & 0.1 & 1 & 0.5 & 0.015 & 0.025 & & as N19, with $N_{\rm SF,neighb}=32$ \\
{\bf N\,21} & 2 & 0.1 & 1 & 0.5 & 0.015 & 0.025 & & as N19, with dynamic softening\\
            &   &     &   &     &       &       & & in the gas\\
{\bf N\,22} & 2 & 0.1 & 1 & 0.5 & 0.015 & 0.025 & isothermal & as N21  \\ 
\hline
 
{\bf N\,23} & 2 & 0.01 & 1 & 0.5 & 0.015 & 0.025 & & as N21, with $\epsilon_{\rm SF}=0.01$\\
{\bf N\,24} & 2 & 0.01 & 1 & 0.5 & 0.015 & 0.025 & isothermal & as N22, with $\epsilon_{\rm SF} = 0.01$\\
{\bf N\,25} & 2 & 0.05 & 1 & 0.5 & 0.015 & 0.025 & & as N21, with $\epsilon_{\rm SF} = 0.05$\\
{\bf N\,26} & 2 & 0.05 & 1 & 0.5 & 0.015 & 0.025 & isothermal & as N25  \\
\hline
 
{\bf N\,27} & 2 & 0.3 & 1 & 0.05 & 0.015 & 0.015 & & as N14, with $\alpha_{\rm crit} = 0.05$\\
\hline
{\bf N\,28} & 2 & 0.3 & 10 & 0.5 & 0.015 & 0.050 & & as N3 \\
\hline

{\bf N\,41} & 2 & 0.3 & 10 & 0.5 & 0.015 & 0.015 & & as N3, with WMAP3 baryon\\ 
            &   &     &   &     &       &        & & fraction\\
\hline

{\bf N\,42} & 2 & 0.3 & 10 & 0.5 & 0.015 & 0.015 & & as N3, with 5 times more\\
            &   &     &   &     &       &        & & baryon and DM particles\\
\enddata
\label{table:models}
\tablecomments{Columns: (1) model (see text); (2) estimated collapse redhift for
the DM halo; (3) the fraction of thermalized energy from the stellar feedback; (4)
the collapse time for SF clouds in units of the free-fall time; (5) fudge factor
for the critical density when triggering the SF; (6) stellar gravitational
softening; (7) limiting value of dynamic gravitational softening in the gas; (8)
notes}
\end{deluxetable*}
The initial density
profile corresponds to the average density around a $2\sigma$ peak in a
Gaussian random density field with power spectrum 
$P(k) = A k^{-2} \exp\left(-k^2 R_{\rm f}^2\right)$. The simulations start at
$z_{\rm i}=36$ to ensure they begin in the linear regime.
The particles are initially moving outward with the Hubble flow, with 
consecutive shells of mass stopping, turning around and falling back
inward.  The model is parameterized by the 
filter mass $M_{\rm f}$, which defines the mass contained within a sphere
of radius $2\,R_{\rm f}$ (e.g., Thoul \& Weinberg 1995). 

The parameters $M_{\rm f}$ and $R_{\rm f}$ are determined by choosing the collapse 
redshift $z_{\rm c}$  at which the shell $R_{\rm f}$ reaches the center
in a non-dissipative collapse and a circular 
velocity $v_{\rm c}$ associated with the shell surrounding the characteristic 
filter mass $M_{\rm f}$ if virialized at half its turn-around radius.
In addition, the model is characterized by the virial coefficient
$q=2T_{\rm rnd}/\left|U\right|$, where $T_{\rm rnd}$ is
the kinetic energy in random motions and $U$ is the potential energy,
and a spin parameter $\lambda=J\left|E\right|^{1/2}G^{-1}M^{-5/2}$,
where $J$, $E$, and $M$ are respectively the total angular momentum,
energy, and mass of the system. 

The initial angular velocity is taken as $\omega \propto r^{-1}$, where $r$
is the cylindrical radius and the central kpc has been softened. The angular
velocity has been normalized by the requirement that the initial $\lambda = 0.05$ 
for all models. This distribution of the specific angular momentum complies
with the universal profile of angular momentum in DM halos, based on the 
statistical sample of halos drawn from high-resolution numerical simulations
(Bullock et al. 2001; see also Fall \& Efstathiou 1980). 

For all the  models 
we adopt the values of $z_{\rm c}=2$, $q=0.01$ 
and $\lambda=0.05$. In addition, we have run models with $z_{\rm c}=1$ 
and 6, but those are not shown here. 
The initial models consists of collisionless DM particles with the total mass
of $7.03\times 10^{11}~{\rm M_\odot}$ and of the gas,
given by the SPH particles, which have the same radial profiles as the DM.
The initial gas comprises 10\% of the total mass.  

\section{Results: Model Evolution}

Our results are presented in this and next sections.
We describe the evolution starting with QN3, the pure DM model, and continue
with models which include all components,
as well as the SF and the feedback from stellar evolution --- 
the standard model N3 is followed by models from Table~1. Each subsection
emphasizes a different aspect of evolution. 
We provide a more specific comparison between models, based on parameter 
variation, in Section~4. Note that the specific times characterizing
the model evolution are given for comparison only.

\subsection{Pure DM Model}

Evolution of pure DM halos within the framework of a monolithic collapse
is limited to a virialization and incomplete violent relaxation in the varying
background gravitational field. We find that the angular 
momentum $J$ incorporated in the initial conditions is essentially
channeled into the internal circulation.
In all models, including those with baryons, the halo forms a prolate triaxial
figure and its minor axis is directed along the $J$-axis. 
The halo {\it figure} tumbles with a pattern speed of 
$\Omega_{\rm h}\sim 1.4~{\rm km~s^{-1}~kpc^{-1}}$ during
the collapse period of $t < 1.9$~Gyr and $\sim 0.2~{\rm km~s^{-1}~kpc^{-1}}$
thereafter. This is more than two orders of magnitude lower than the pattern 
speeds of other prolate figures, i.e., stellar or gaseous bars, forming within 
the growing disk, even {\it after} the bar secular slowdown. Overall the halo 
figure makes $\ltorder 200^\circ$, or less than one full rotation, in a Hubble 
time. Hence, for all practical reasons, we can consider that the halo figure is 
static with respect to any developing bars in the system. The initial 
$\lambda=0.05$ parameter remains constant for QN3 and 
shows an increase by $\sim 30\%$ for models with gas, over the Hubble time.
The DM density profile in QN3 quickly achieves the NFW shape and the
characteristic radius $R_{\rm s}$ is $\sim 9$~kpc by the end of the run. 

Next, we determine the evolution of the DM halo shape in QN3. The inner 30~kpc
halo turns around first and develops a triaxial structure between $\sim 0.5-1$~Gyr.
Figs.~1$a,c$ show the evolution of axial ratios, $b/a$ and $c/a$, in QN3. 
Beyond  $\sim 50$~kpc, the DM collapses over $\ltorder 5$~Gyr 
and forms a strongly triaxial figure as well. The initial triaxiality develops in 
response to a short-lived 
radial orbit instability (e.g., MacMillan et al. 2006). 
The DM halo in QN3 exhibits a clear increase in its triaxiality with
radius, from the inner, $\ltorder 20$~kpc, to the outer, $\gtorder 40$~kpc, halo,
both in its prolateness and flatness. The former range is $0.1 \ltorder 
\epsilon_{\rho} \ltorder 0.4$, while the latter lies within $0.35 \ltorder 
f_{\rho} \ltorder 0.55$. Both parameters show a substantial increase into the 
outer halo. 

Roughly speaking, the inner halo is assembled during the first Gyr while 
the outer one during $\sim 5$~Gyr. 
Fig.~2 shows the growth of the halo virial mass, $M_{\rm vir}$ which is 
computed by requiring that the mean halo density is equal
to the critical density of the universe times $\Delta_{\rm c}$, where  
$\Delta_{\rm c}$ is obtained from the top-hat model (e.g., Romano-Diaz et al. 
2006, 2007). The halo virial mass reaches $7.4\times 10^{11}~{\rm M_\odot}$ by
$z=0$. The halo growth rate agrees well with the fully cosmological simulations 
in OCDM universe (e.g., Romano-Diaz et al. 2007). Unlike Samland \& Gerhard 
(2003), we do not assume the law of halo growth and do not limit the shape of the 
halo to be spherically-symmetric.
\begin{figure}
\begin{center}
\includegraphics[angle=-90,scale=0.47]{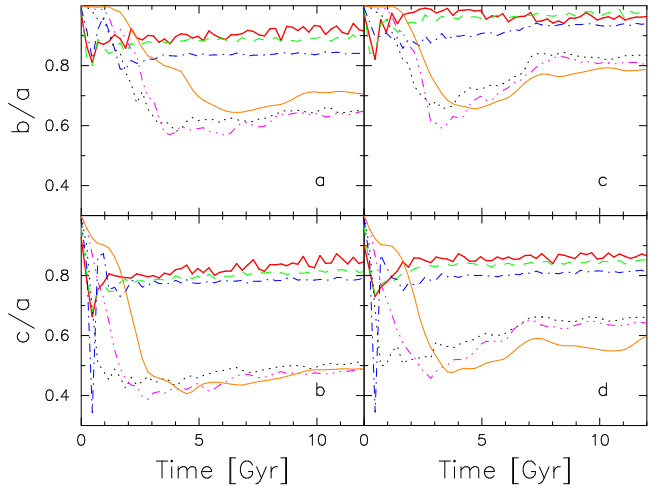}
\end{center}
\caption{Evolution of the axial ratios $b/a$ ($a$ and $c$) and $c/a$ ($b$ and $d$)
for pure DM halo model, QN3 and for the standard model N3. The curves
correspond to 2~kpc (solid red), 5~kpc (dashed green), 10~kpc (dot-dashed blue),
50~kpc (solid orange), 100~kpc (dotted black) and 300~kpc (dash-dot-dot-dotted
magenta).
\label{fig.1}}
\end{figure}

\begin{figure}
\begin{center}
\includegraphics[angle=-90,scale=0.4]{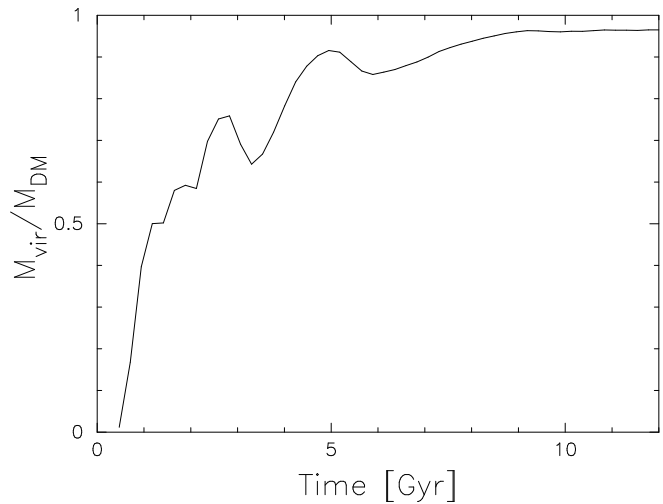}
\end{center}
\caption{QN3 model: assembling the DM halo within its virial radius. Shown is 
$M_{\rm vir}/M_{\rm DM}$ ratio, i.e., the virial mass normalized by the total
DM in the simulation. The inflow rate is $\sim 600~{\rm M_\odot~yr^{-1}}$
during the first $\sim 1$~Gyr and $\sim 15~{\rm M_\odot~yr^{-1}}$ averaged over 
the next 9~Gyr.}
\end{figure}

\subsection{Evolution of the Standard Model and Beyond}

Next, we present models that include baryons and their associated 
processes. The standard model, N3, is treated as a representative one, and
is compared to other models. The $\lambda$ parameter is
growing from 0.05 by $\sim 30\%$ in all of these models. We note that {\it the 
very slow tumbling of the halo figure is not affected by the presence 
of the baryonic matter} --- the angular momentum is channeled mainly into the 
internal circulation. 

The overall trend in the evolution of a DM halo remains the same in the presence 
of baryons (Fig.~1$a$-$d$). but its prolateness and flatness  
are lessened by $\sim 10\%-20\%$ compared to that in QN3. This means that the  
outer halo stays substantially more prolate and flattened than the inner one, as in QN3.
The maximal prolateness and flatness of the outer halo are reached earlier in N3
and its subsequent reduction is stronger there over the next 2--3~Gyr.
The inner 10~kpc also become more axisymmetric than in QN3. Asymptotically, N3 
is substantially less prolate than QN3. The 
washing out of the prolateness in the inner 20~kpc is clearly connected with the 
buildup of the disk over the first 2--3~Gyr (as shown already by Berentzen \& 
Shlosman 2006). 

Two factors play the major role in removing the prolatenes of the inner halo.
First, the baryonic (density) response to the halo (equatorial) prolateness
is perpendicular to the halo's nearly non-rotating major axis. In this case
the inner inner Lindblad resonance (inner ILR), if it exists, moves to the center, 
while the outer ILR to the very large radii, outside the halo, 
securing this response everywhere within the halo. This baryonic response dilutes 
the prolate potential of the DM (more about this response in section~3.6). Second, 
the disk sitting deep inside the DM
potential, as well as numerous baryonic clumps forming {\it prior} to the disk assembly,
scatter the DM particle orbits, further reducing its triaxiality.
The gravitational quadrupole of the disk, during its assembly, interacts with that 
of the prolate
halo --- this induces a chaotic behavior in the halo and in the disk. The result
is a reduced prolateness and flatness in the halo and a more axisymmetric disk.

For the stellar component, the inner $\sim 5$~kpc of N3 show a mild prolateness 
asymptotically, with the axial ratio of $b/a\gtorder 0.95$ at $\sim 10$~kpc due
to a mild bar. The ability of the DM to maintain
the non-negligible prolateness affects both the dynamical evolution of the galactic 
disk and facilitates the angular momentum redistribution in the disk-halo system,
as we discuss in section~3.8.

Because dissipative processes dominate the formation of a stellar disk,
it is dramatically flatter than the DM. For N3 and
other models with baryons discussed here, the disk formation and 
evolution can be divided roughly into two stages: gas-dominated and star-dominated. 
The former stage lasts over the first 2--3~Gyr --- we observe that the gas dominates
over the stars even in the central regions of the disk. The latter stage lasts over 
the rest of the simulation time,  during which the ratio of 
baryonic-to-DM matter within the central 10~kpc reaches $\sim 0.6-0.65$ and 
subsequently rises to $\sim 0.74$. The inner few kpc, however, are baryon dominated. 
The gas rotation curve flattens at about 290~${\rm km~s^{-1}}$.  The 
vertical thickness of the gas layer is large over the first 5~Gyrs and then is 
gradually reduced.

The initially gas-dominated disks are not axisymmetric, and exhibit a range of
morphologies: the $m=2$ and $m=3$ modes dominate them at various times, and
occasionally they are affected by the off-center, $m=1$ mode. With time, the disks 
become more axisymmetric, but some stay oval for a long period of time. 
\begin{figure*}
\begin{center}
\includegraphics[width=6.0cm,angle=0]{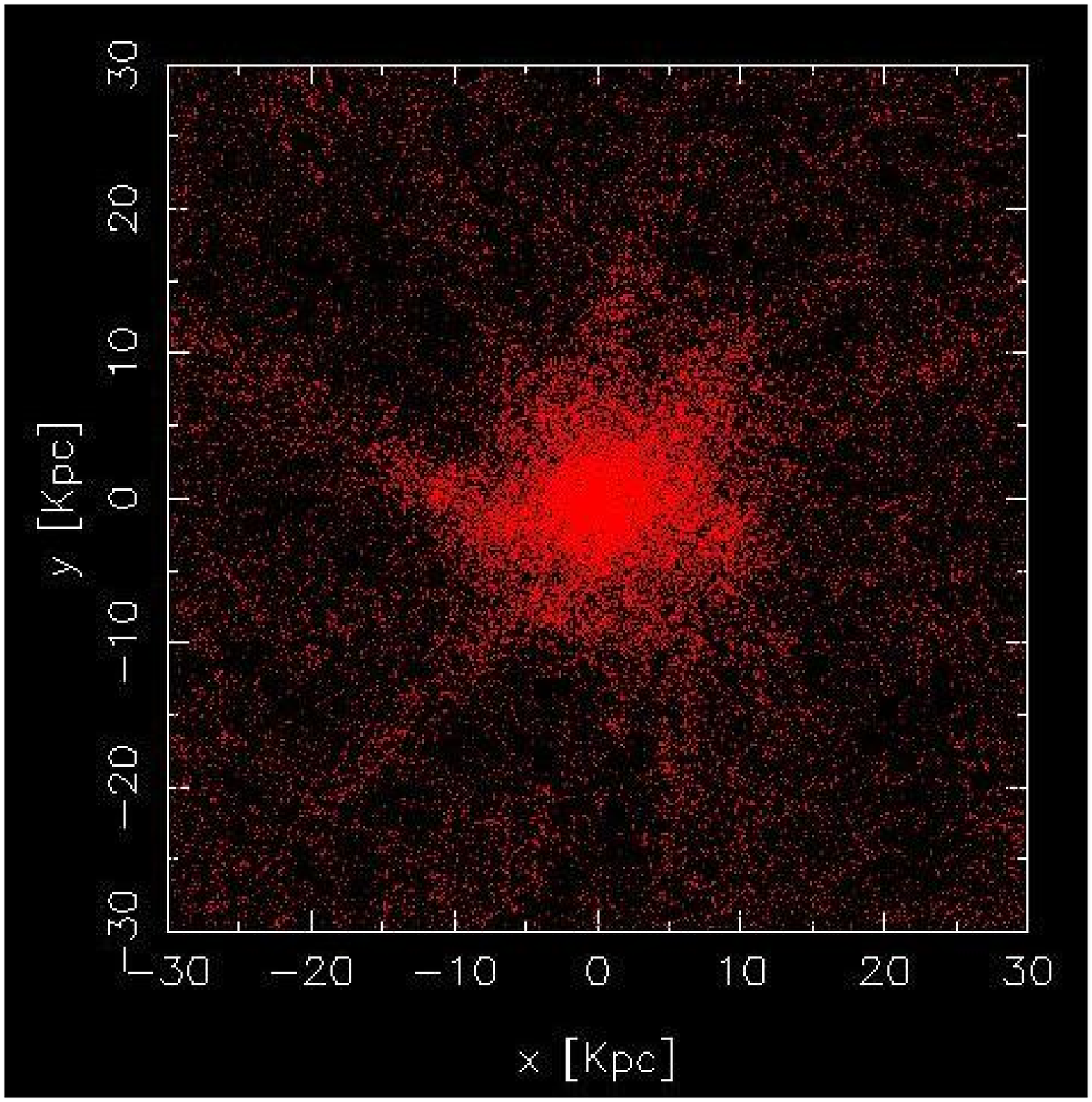}\hspace{0.2cm}
\includegraphics[width=6.0cm,angle=0]{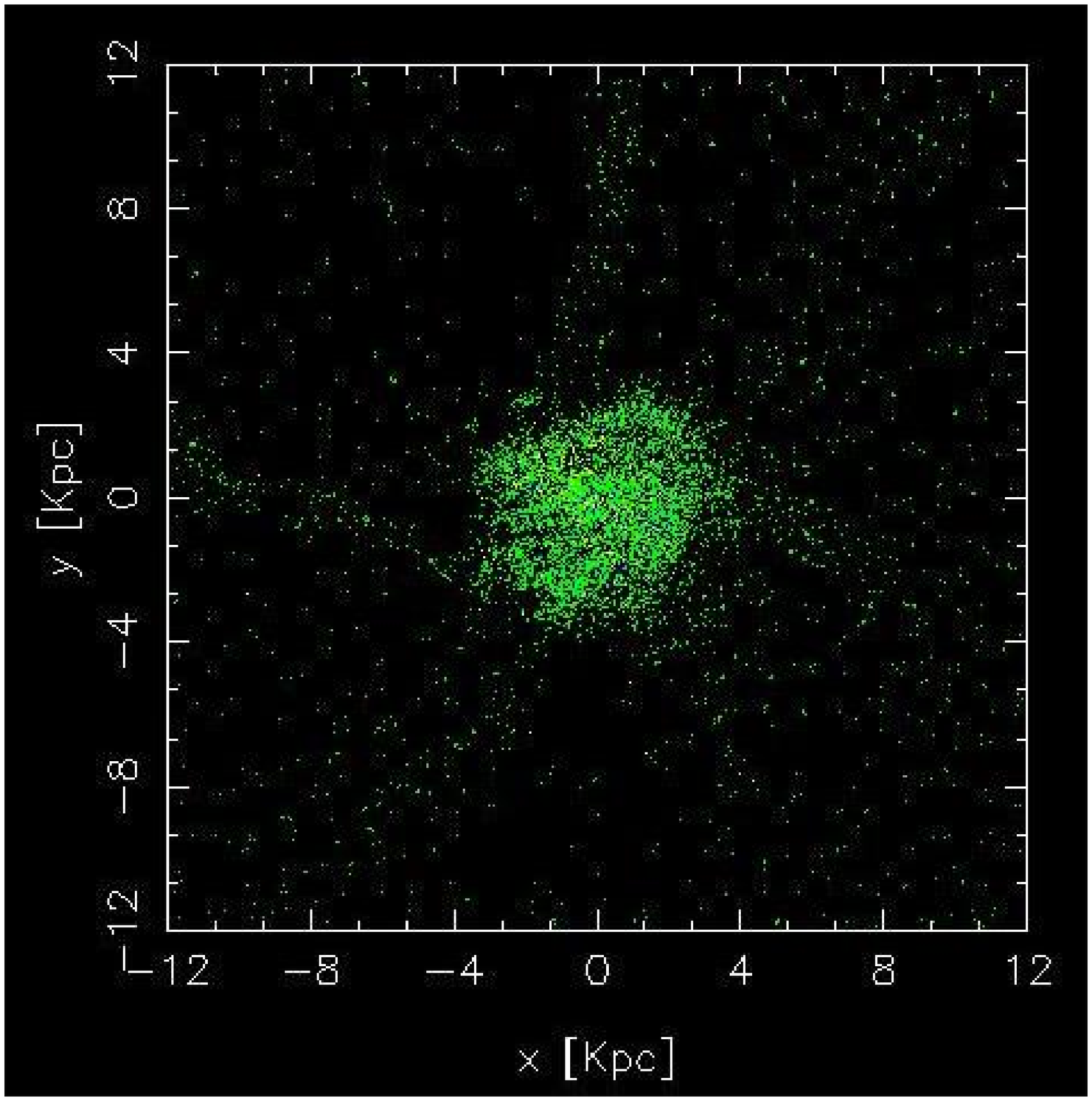}\\
\end{center}
\caption{Snapshots of inner DM halo (left) and disk (right) formation in the N3 model.
The time, $\tau=0.71$~Gyr was selected to emphasize the `cat's cradle' morphology
of the proto-disk which displays the attached radial filaments. This morphology
is driven by an underlying DM potential of the forming inner halo. The disk
particles consist of gas (green), SF particles (blue) and stars (yellow). Left frame 
size is $30\times 30$~kpc and right frame is $12\times 12$~kpc}
\end{figure*}

Even
towards the end of the simulation in N3, only a fraction of baryons resides within the
central 10~kpc (where most of the disk lies): $5.8\times 10^{10}~{\rm M_\odot}$ 
in stars and $5\times 10^{9}~{\rm M_\odot}$ in gas. The total mass of baryons within
this radius is $6.3\times 10^{10}~{\rm M_\odot}$, or 49\% of the baryons in the 
simulation. A similar baryon fraction, $\sim 40\%-60\%$, is characteristic for all models. 
Between a third and a half 
of all baryons reside outside the SF region in the disk plane (beyond 15~kpc) and in 
the DM halo, i.e., in the form of partially rotationally-supported and/or hot gas.
The rest of the gas is shock-heated to virial temperatures of $\sim 10^6$~K, and ionized.

The DM density profile for N3 can be fitted by the NFW law  although the quality of the
fit is lower than that of QN3, and the final value of $R_{\rm s}$ is $\sim 4.4$~kpc.
The inner few kpc and the DM cusp are dominated by the
baryons at the end of all the runs with the gas.

\subsection{Early Baryon Inflow and the Disk Formation:\\ the Cat's Cradle}

Early baryon inflow towards the center leads to the formation of a specific 
configuration near or at the DM density peak that resembles the ``cat's 
cradle'' pulled by a number of nearly radial `strings' (filaments). This is observed in 
all the models, except in N4 which has an enforced
axisymmetric halo. The source of this morphology is the background DM
distribution within the inner $\sim 30$~kpc halo during the initial $\sim 1$~Gyr
(Fig.~3, left). The baryonic strings have one-to-one correspondence with
the DM strings. Furthermore, the face-on disks show such strings attached to their 
edge, where the main baryonic inflow joins the disk. The strings are not perfectly
radial, and at the disk edge are tangent to it --- {\it the cold gas inflow, therefore,
joins the disk without being shocked, thus supplying a cold, unvirialized gas to the
disk}. The typical symmetry resembles the irregular polygons, and ranges 
from triangles to octagons (e.g., Fig.~3, right). The central object --- an amorphous clumpy 
protodisk, has a typical size of 2--4~kpc and the radial 
strings attached. The protodisk has an elongated shape and its figure initially nearly 
stagnates. The baryonic blobs tend to form on these strings in the equatorial 
plane and fall along them onto the disk, visibly speeding up its figure rotation. The 
feeding of the disk along the
strings continues for some time during the first Gyr. Models N23--N26
with a small SF feedback, $\epsilon_{\rm SF}=0.01$, exhibit larger clumps. 
The early disk appears as a fat configuration with an equatorial SF and with flaring
edges. By about $\sim 1$~Gyr, the disk grows substantially, its shape
changes rapidly between a less and a more prolate one, and with more or less prominent
spiral structure. Dynamical friction appears to play a major role following the clump
joining the disk --- similar to the behavior of the gas-rich disks 
analyzed by Shlosman \& Noguchi (1993). 

\subsection{Later Stage: Baryon Mass Inflow History and the Disk Growth}

With the beginning of the collapse, the rate of baryonic inflow into the
inner 20~kpc reaches its maximal rate within the first 1.5~Gyr and 
then steadily falls off. The SF rate shows a similar behavior (except in N27), with 
a peak at 1.5--1.8~Gyr of $\sim 38~{\rm M_\odot~yr^{-1}}$ for the N3 disk (Fig.~4). 
Overall, the range of the SF rates is $10-60~{\rm M_\odot~yr^{-1}}$, and the
peak SF rate can be delayed by 1--2~Gyr compared to the standard model. N27
differs and shows a flat SF rate for the first 6---7~Gyr.

\begin{figure}
\begin{center}
\includegraphics[angle=-90,scale=0.4]{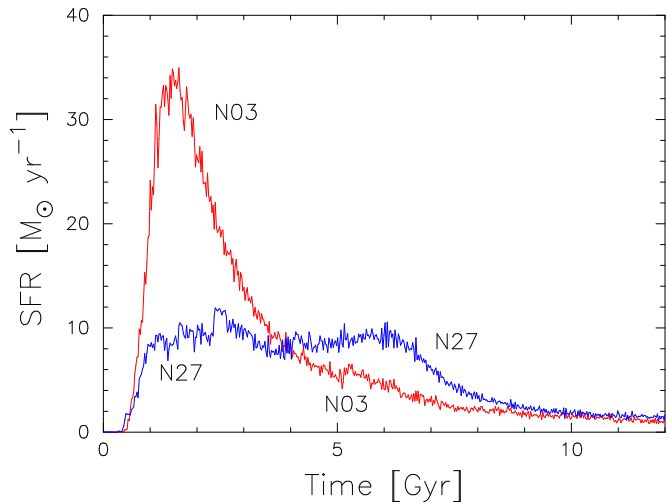}
\end{center}
\caption{Star formation rates in the N3 and N27 model disks. }
\end{figure}
\begin{figure*}
\begin{center}
\includegraphics[width=4.2cm,angle=-90]{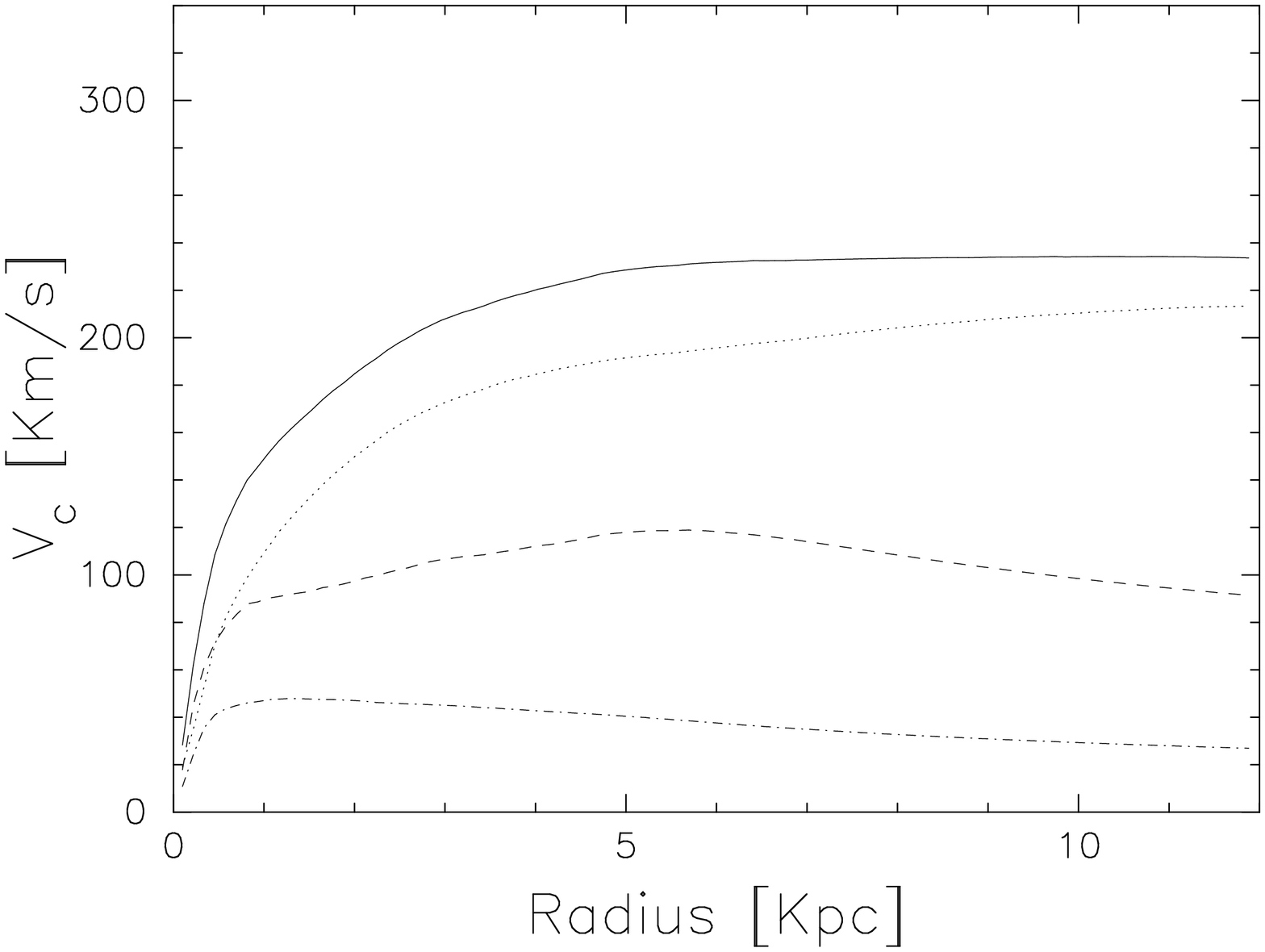}\hspace{0.1cm}
\includegraphics[width=4.2cm,angle=-90]{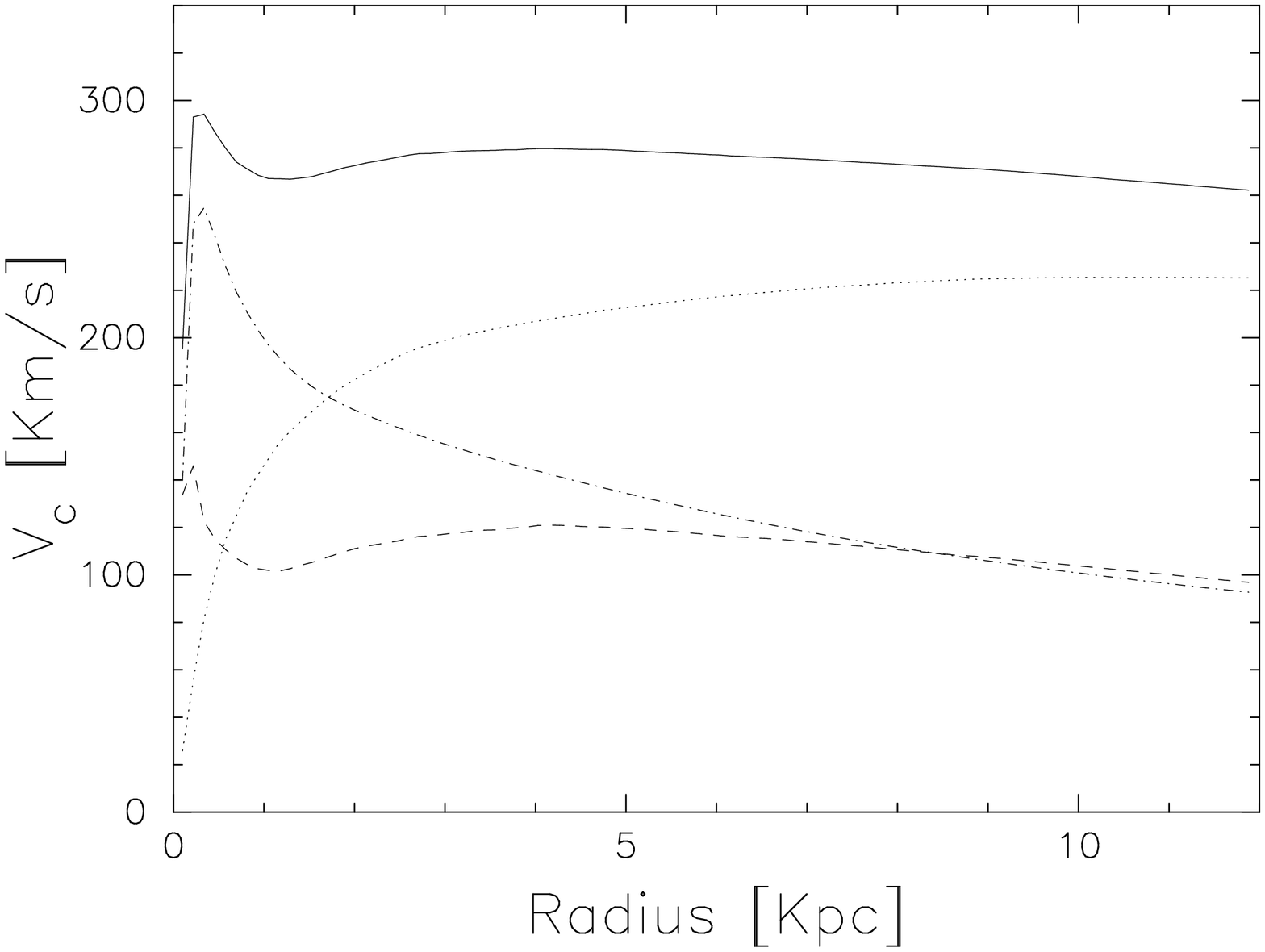}\hspace{0.1cm}
\includegraphics[width=4.2cm,angle=-90]{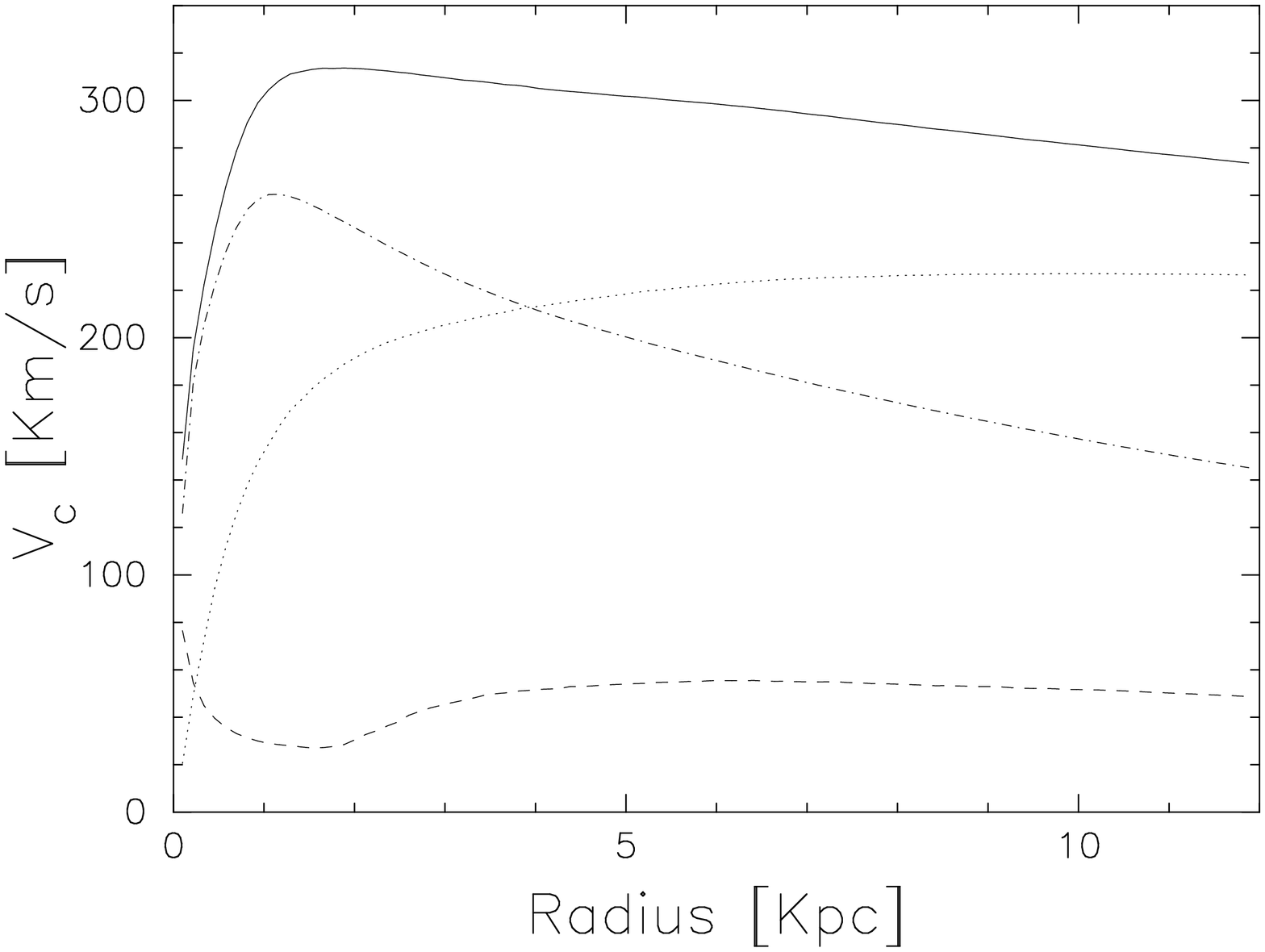}\\
\end{center}
\caption{Snapshots of a circular velocities in the assembling disk/halo system of model
N3 at 1~Gyr (left), 2~Gyr (center) and 12~Gyr (right). Total 
velocity is given by the solid line, DM by dotted line, stellar by dash-dotted, and
gaseous by dashed lane.}
\end{figure*}

The SF rates decay exponentially, in nearly all models (except N27), to $\sim 
5~{\rm M_\odot~yr^{-1}}$, at $\sim 5$~Gyr, and $\sim 1~{\rm M_\odot~yr^{-1}}$
after $\sim 10$~Gyr. The latter rate is typically observed in the normal disk galaxies
of the nearby Universe. 

In all models, the proto-disk plane has been established and visible already at 
$\tau\sim 0.5$~Gyr, across the inner 12~kpc, while the stars exist 
only at the very center by that time. Thereafter, a small 
$r\sim 2$~kpc gas disk becomes prominent and doubles in size to 5~kpc in another
0.5~Gyr, with the SF at its midplane.  After $\sim 1$~Gyr, a gaseous bar of 
$r\sim 2$~kpc forms, supplemented by an ongoing SF, but collapses to the center and 
is damped shortly (more about these bars in sections~3.5--3.7).   
By $\tau\sim 1.2$~Gyr the disks grow almost to 10~kpc with an intense SF --- the
gas component remains puffed up significantly.  

By $\tau\sim 2$~Gyr, the gas layer starts to cool down.
The transient spiral structure becomes visible and is delineated by the SF regions. 
By $\sim 6$~Gyr, the vertical structure of the stellar disks in most of the models,
but not in all, looks alike. We observe that the {\it disk cools down and thins 
substantially when the SF rate drops below
$\sim 5~{\rm M_\odot~yr^{-1}}$ --- this is true for all the models, and typically 
happens after about 5~Gyr.}

All the disks, except that of N27, appear to form from {\it inside out}, and are 
gas-dominated initially. 
The typical time for the disk buildup, i.e., when it reached about 50\% of its
final mass at $z=0$, is $\sim 2$~Gyr. The stellar rotation curve gradually
extends to larger radii and steadily rises with time, while remaining flat in $r$. 
Its peak moves in to smaller radii, then recedes somewhat (Fig.~5).
(See also the comment on the bar pattern speeds evolution, reflecting the growing mass
concentration in the disk, in section~3.6.). The gaseous component, in the inner few
kpc, forms a high surface density `thin' disk embedded in the thicker
`corona.' This corona extends to larger radii.
The extent of the stellar disk is limited to the 
narrow gas disk only. While the relative details of this disk-corona morphology
differ from model to model, the overall structure is robust.   
Models N19 to N26, all show larger disks, some with a strikingly visible bulge 
accompanied by SF activity in the nuclear rings.
 
The central kpc of an assembling galaxy in N3 is dominated by the
DM during the first $\sim 1$~Gyr, and subsequently by the
gas (for a short period of time) --- then by the newly formed stars. 
The baryon/DM ratio tends to 0.8 within the disk radius at the end of the simulation. 
This value, however, is hardly a standard. It is affected by the feedback
from the stellar evolution, by the critical density for SF and by the collapse
time of unstable SF clouds.  

\subsection{Early Nuclear Bars}

\begin{figure*}
\begin{center}
\includegraphics[width=6.cm,angle=-90]{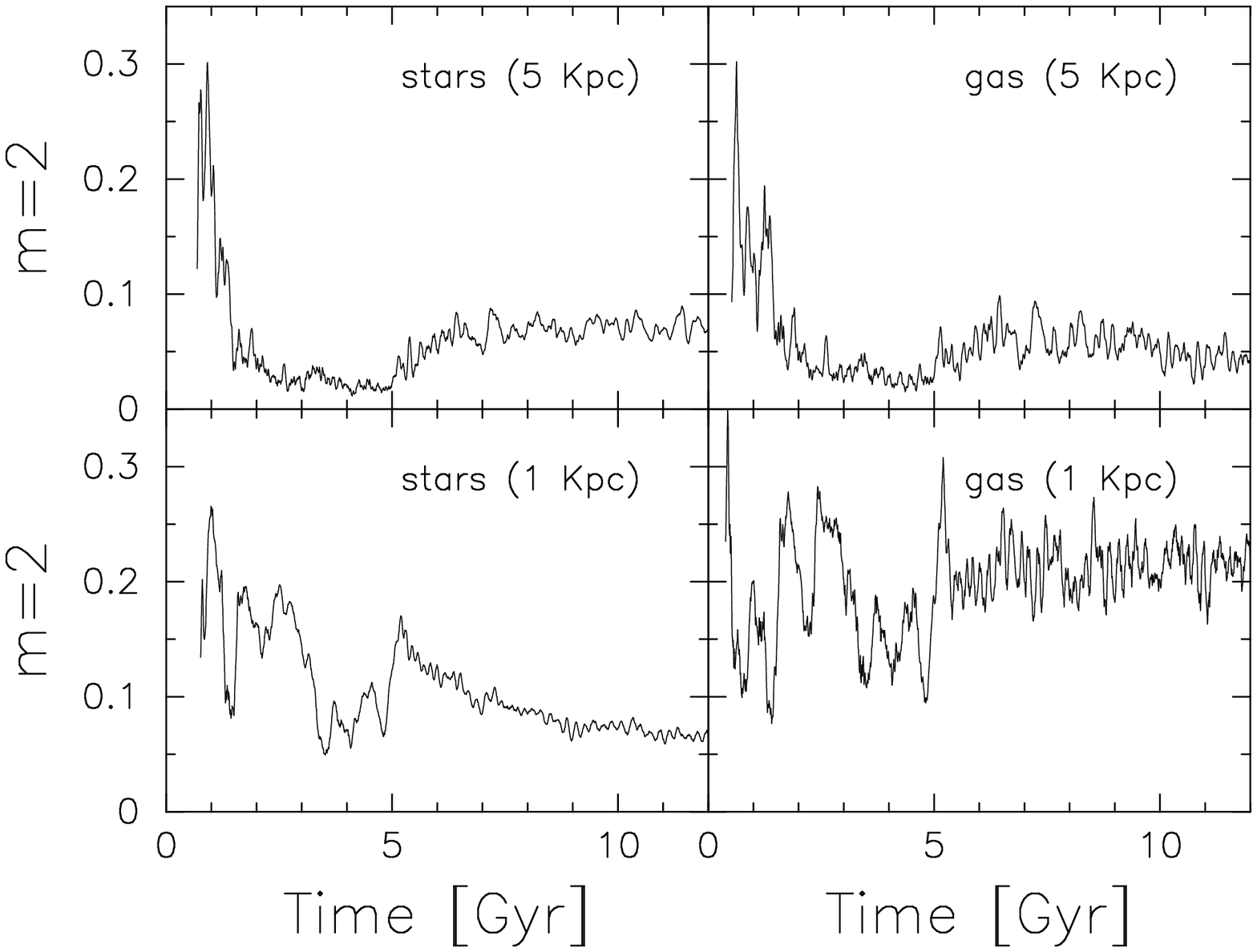}\hspace{0.1cm}
\includegraphics[width=6.cm,angle=-90]{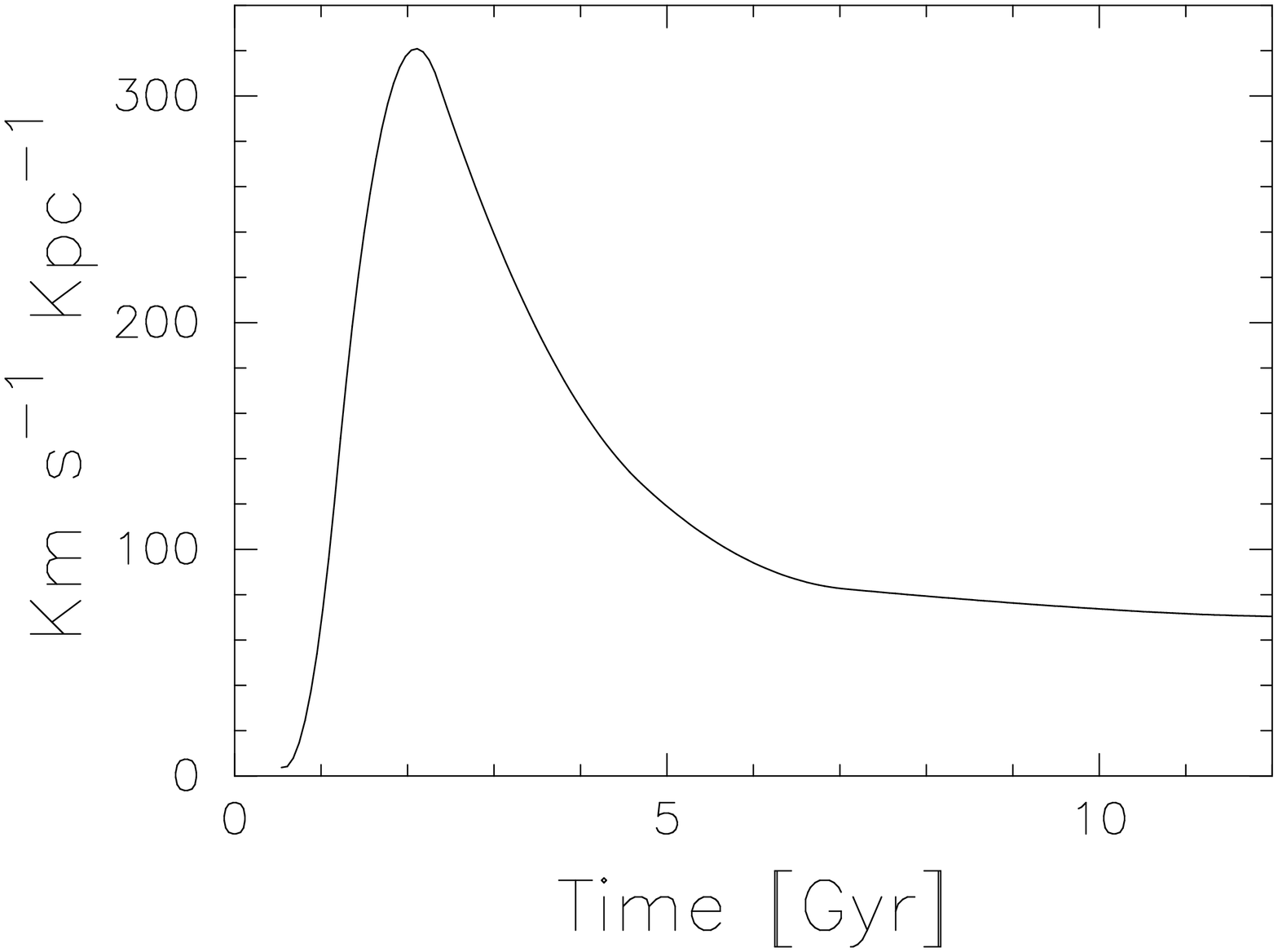}
\end{center}
\caption{\underline{\it Left frame:} The evolution of $m=2$ Fourier component 
$A_2$ in the N3 disk. Shown are stars (left frames) and gas (right) within 
the central 5~kpc (upper frames) and 1~kpc (lower). \underline{\it Right
frame:}
The pattern speed of a stellar bar within the inner kpc of N3
smoothed by fitting a cubic spline to the P.A. of the $m=2$ in Fig.~6$a$
and differentiated. 
\label{fig.9a}}
\end{figure*}
%
%
\begin{figure*}
\begin{center}
\includegraphics[width=5.5cm,angle=0]{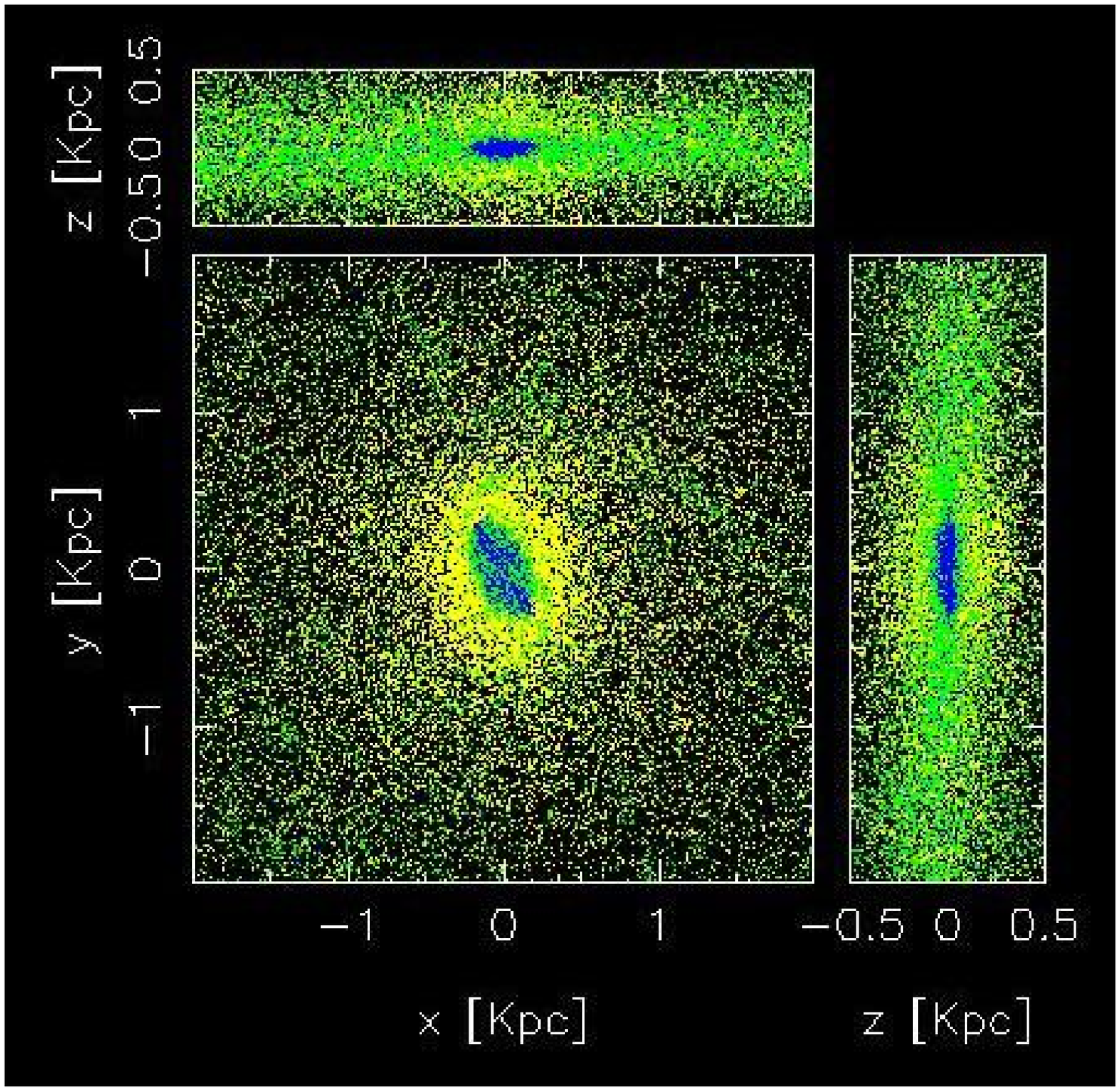}\hspace{0.1cm}
\includegraphics[width=5.5cm,angle=0]{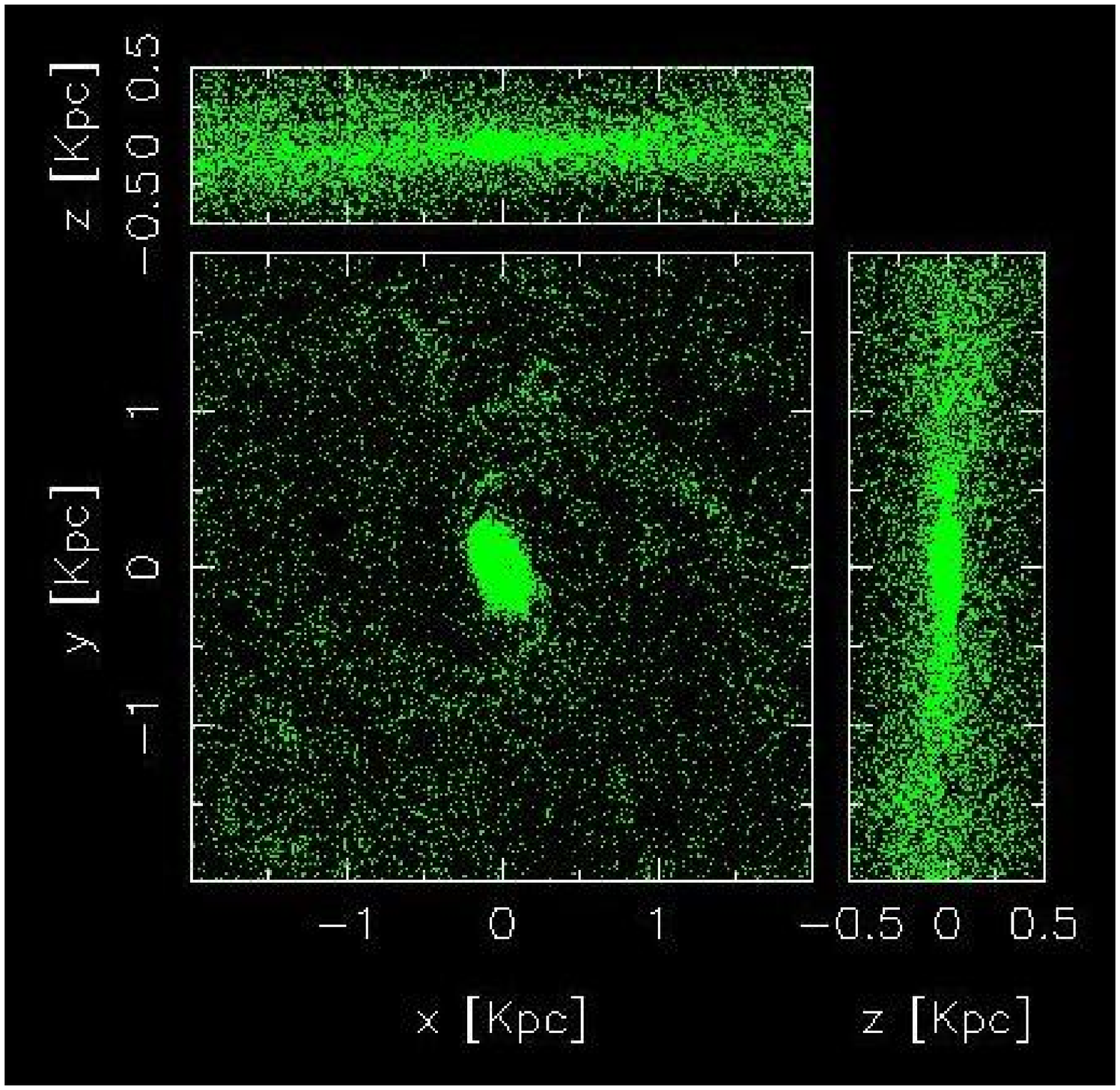}\hspace{0.1cm}
\includegraphics[width=5.5cm,angle=0]{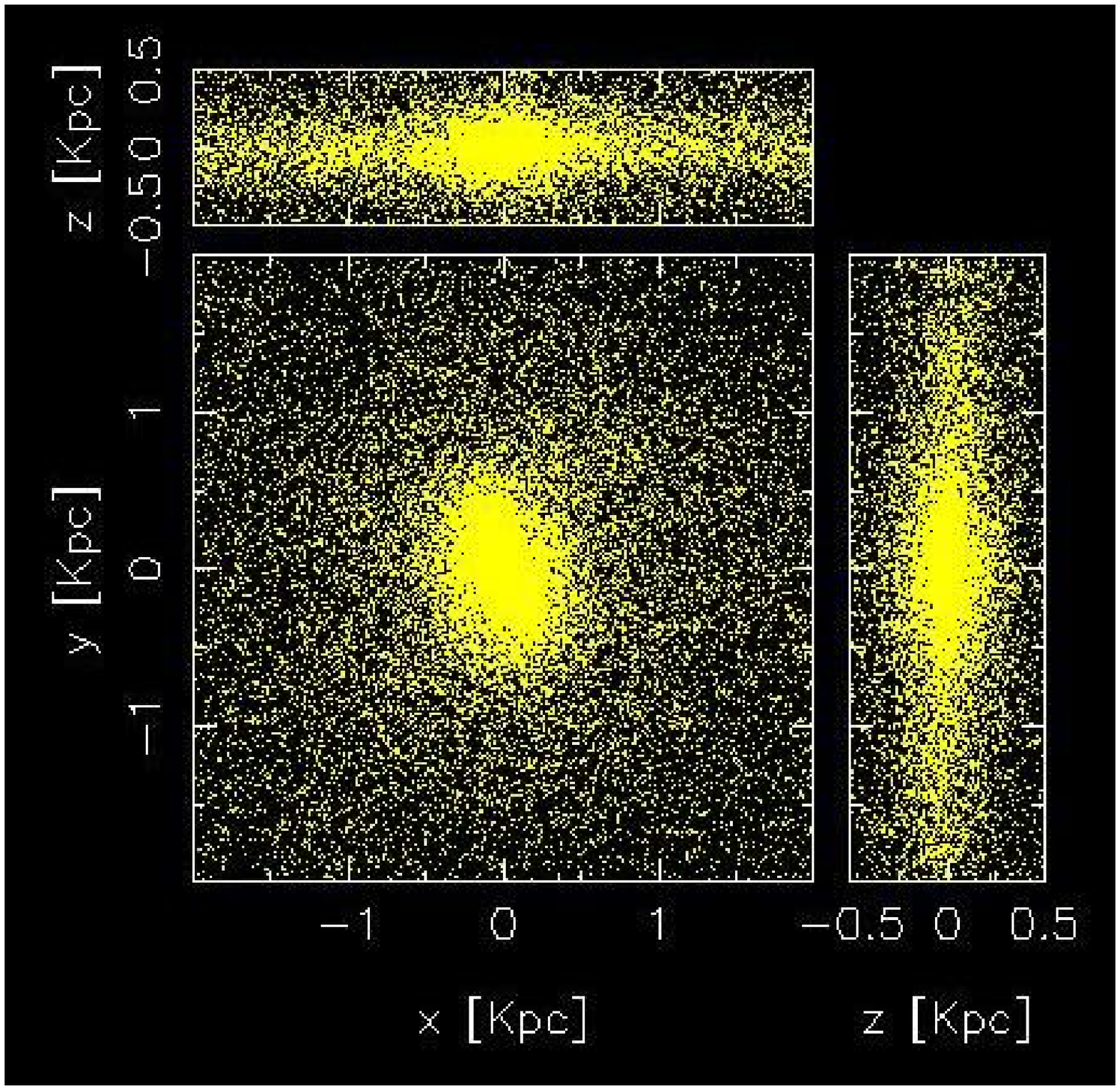}\\
\end{center}
\caption{Example of early nuclear gas bars: N3 bars shown in gas$+$stars$+$SF 
(left frame), gas (center) and stars (right)  at $\tau=2$~Gyr.
}
\end{figure*} 
\begin{figure*}
\begin{center}
\includegraphics[width=5.5cm,angle=0]{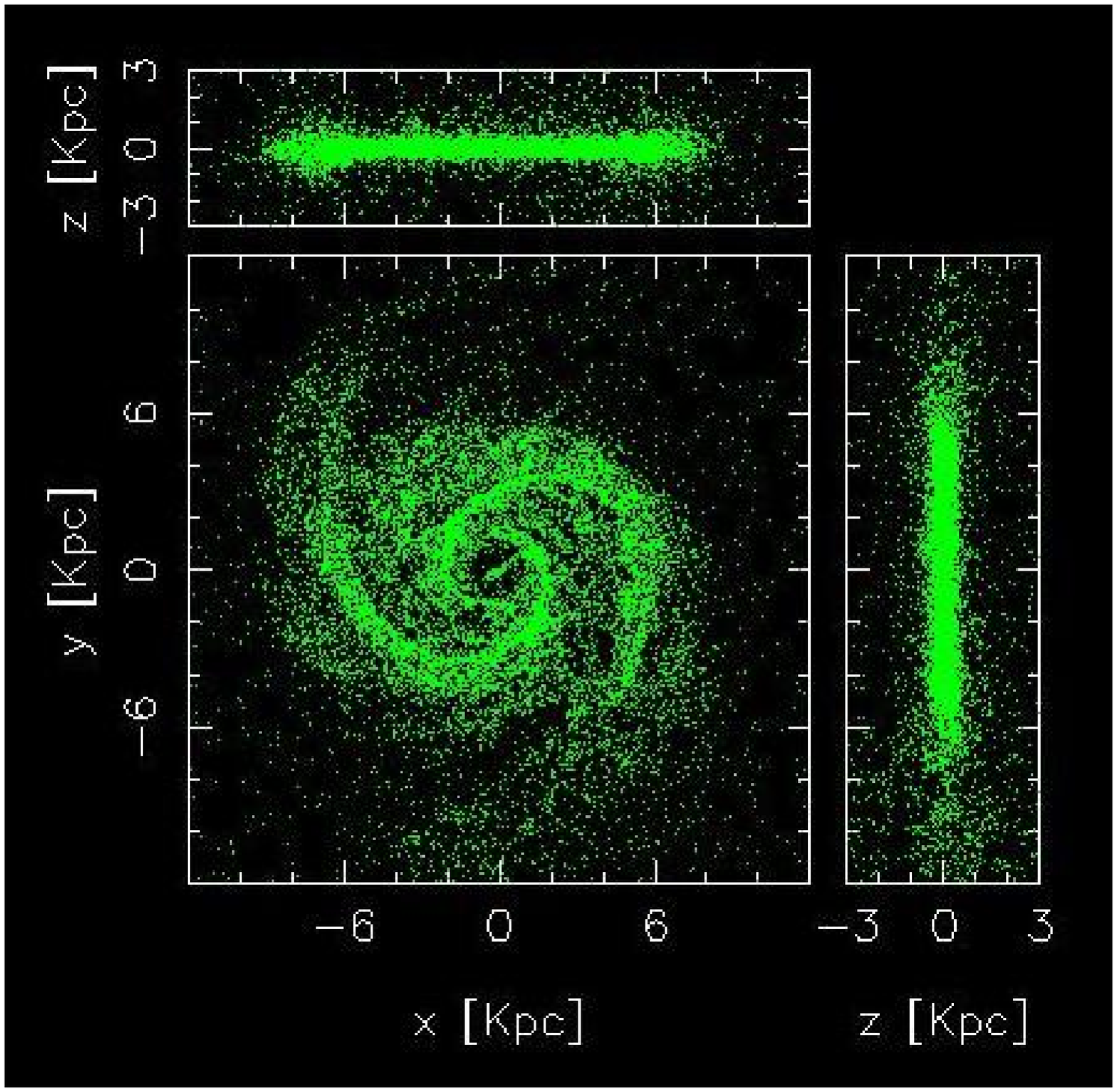}\hspace{0.1cm}
\includegraphics[width=5.5cm,angle=0]{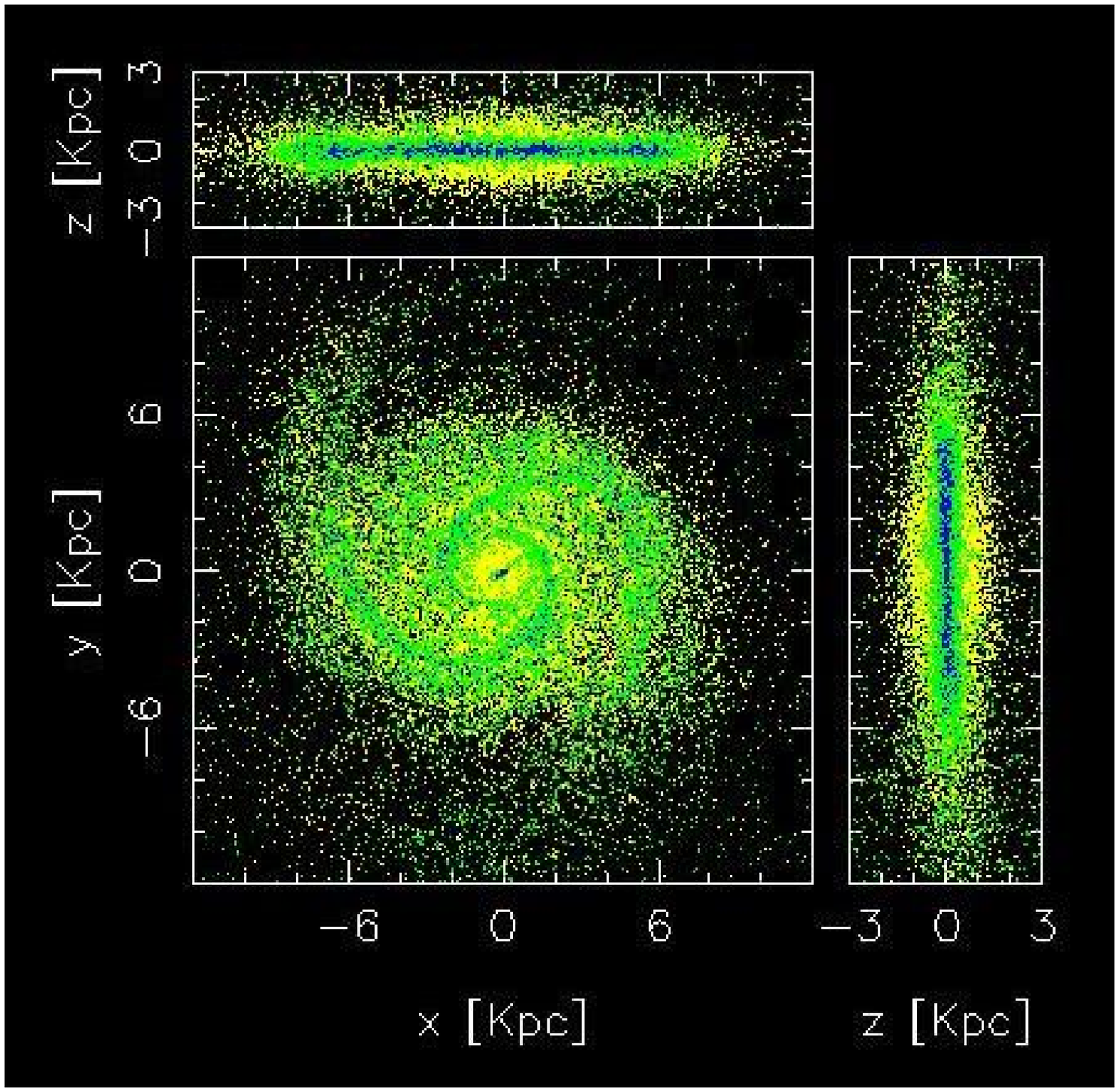}\\
\end{center}
\caption{Example of nuclear rings around nuclear bars: N19 is shown in gas
(left frame), gas$+$stars$+$SF (right) at $\tau\sim 3$~Gyr.
}
\end{figure*} 
\begin{figure*}
\begin{center}
\includegraphics[width=6.0cm,angle=0]{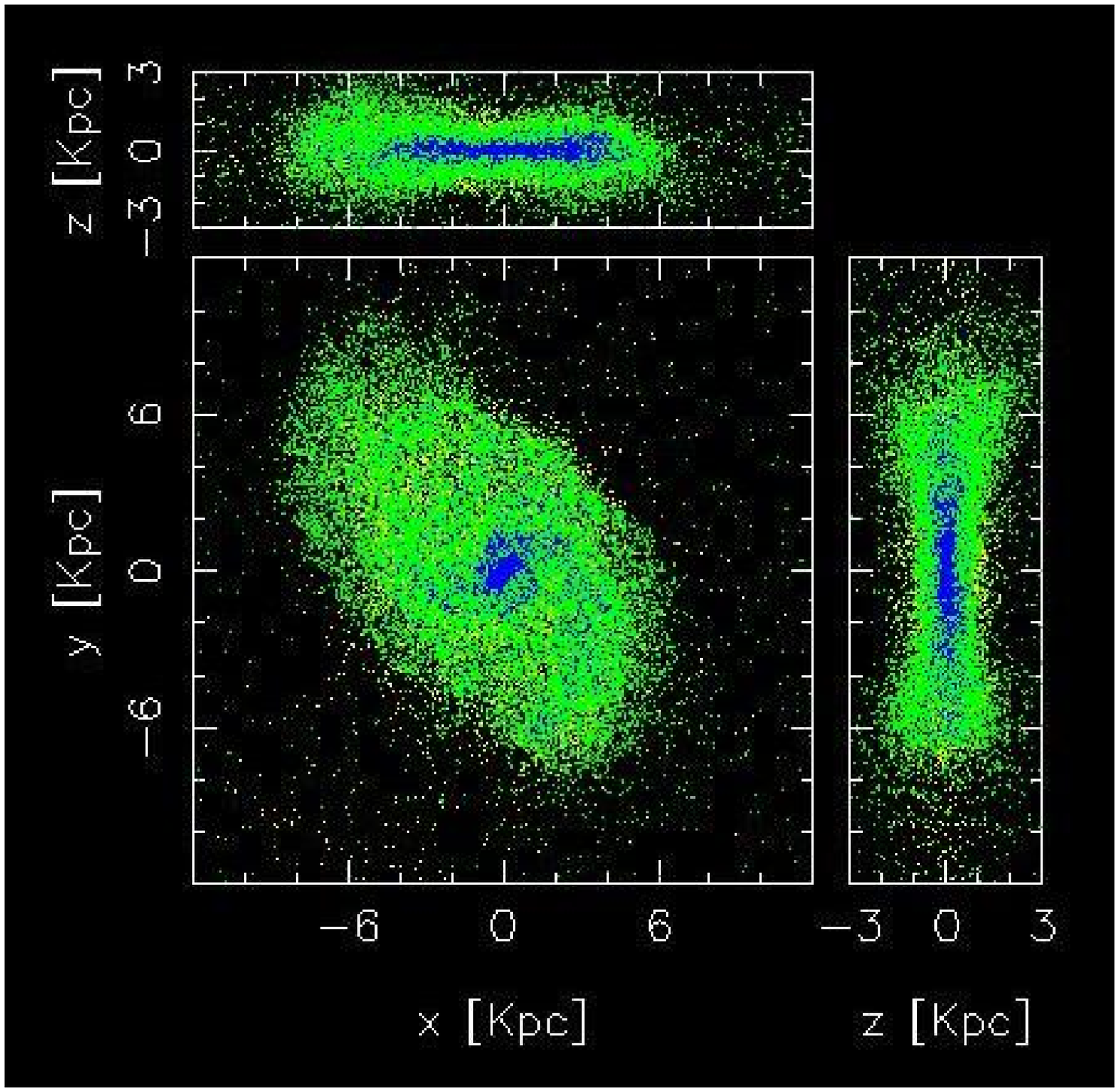}\hspace{0.1cm}
\includegraphics[width=6.0cm,angle=0]{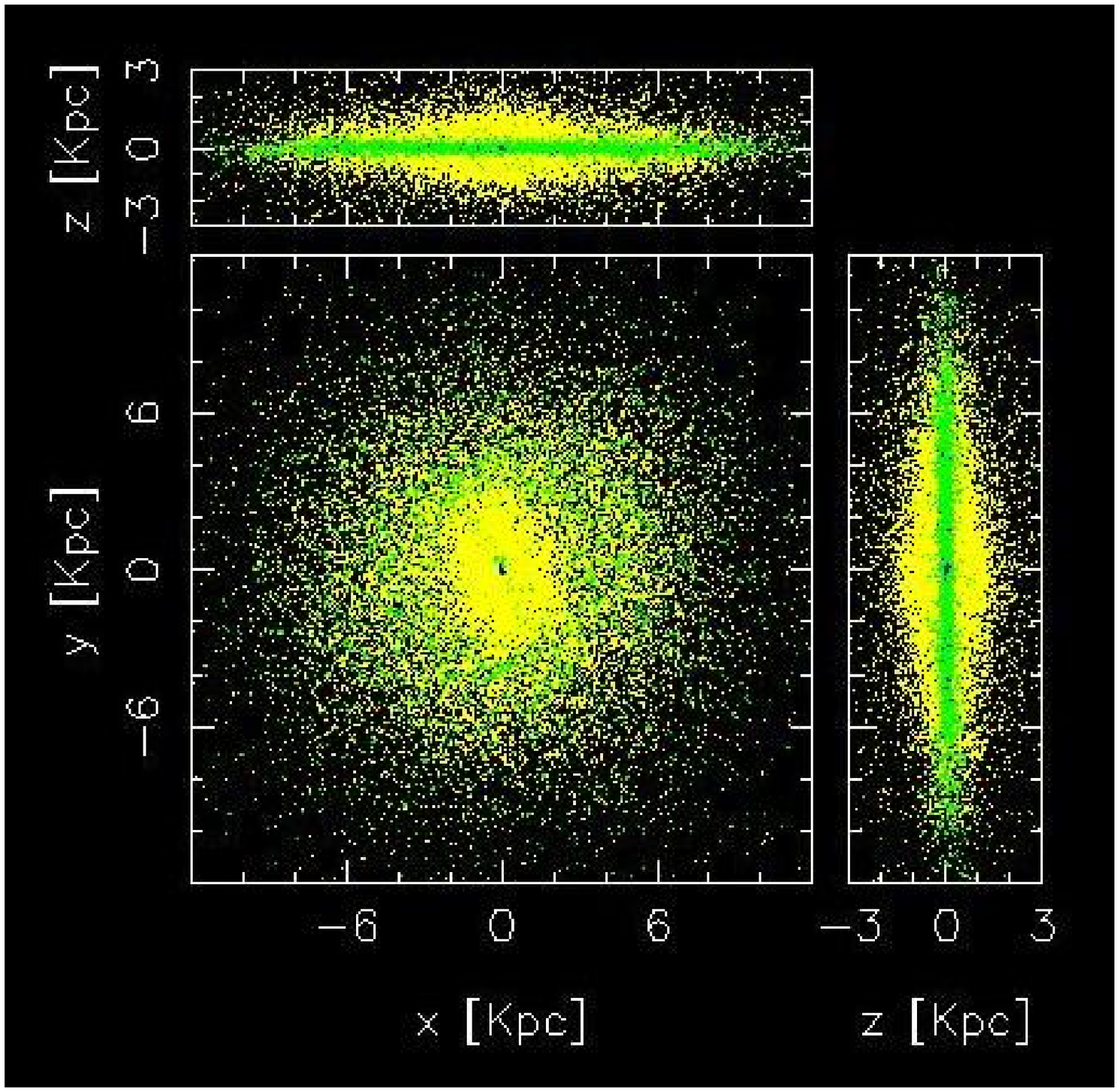}\\
\end{center}
\caption{Examples of early (gas-rich) and late (gas-poor) large-scale bars 
shown in gas$+$stars$+$SF
colors: 8~kpc bar in N5 (left frame) at $\tau\sim 1.5$~Gyr, and 3~kpc bar
in N16 (right frame) at the end of the simulation.
}
\end{figure*} 

Early gas-dominated (or gas-rich) nuclear bars appear as a common feature after
about 1~Gyr in all of the models, albeit their longevity differs from model
to model (typically 4--5~Gyr), e.g., N3 in Figs.~6--7. Their strength, measured by
their axial ratio or by the amplitude of their $m=2$ mode,
changes with time, and they become more star-dominated as the gas 
is channeled inward and 
progressively concentrates in the central $\sim 100$~pc. The (radial) sizes of these
nuclear bars change with time as well, from few 100~pc to about 2~kpc. The 
central SF is clearly concentrated in these bars. The stellar bulges are observed 
to dominate these bars after 2.5--3~Gyr in some models. The vertical extent of the 
gas layers within the bulges appears to be similar to elsewhere in the disk. 
We note that nuclear {\it gas} bars, devoid of the stellar component, are 
observed in the central kpc of our models as well. {\it These frequent gaseous 
nuclear bars survive for a limited
time, depending on the gravitational coupling from the background potential, 
and collapse to the center}.

Nuclear rings appear prominently around the nuclear bars from time to time.
They seem to be connected to the large-scale disk by pair of gas spiral arms
which are also visible in the SF (the outer disks are visibly oval), e.g., N19 in 
Fig.~8. These rings become more prominent when the gas in the oval disk is channeled to 
the central kpc. The gas nuclear bars, seen to develop inside the ring, shrink 
catastrophically and speed up, until the gas is dumped at the very center ---
leaving a profound central hole in the cold gas distribution.

\subsection{Large-Scale Stellar Bars}

Large-scale bars, $\sim 3-6$~kpc, have always developed very early, by 
$\tau\sim 1$~Gyr, e.g., in N5 --- in all cases they have decayed in a few Gyr.
In addition, bars have formed late in the evolution, after 5~Gyr, e.g., in N16 
(Fig.~9), in about 2/3 of the models.
The early bars appear in response to the prolateness in the DM halo 
(and of the stellar disk consequently) --- {\it they do not develop as 
a result of a classical bar instability}. These bars are 
moderately strong, with $\epsilon\sim 0.4-0.5$, and seem to involve a large 
fraction of the stellar disk mass. Their shapes as well as the shapes
of the outer disks depend on the mutual orientation between the DM halo, disk and 
bar major axes. We use the $m=2$ mode amplitude, $A_2$, to quantify the strength
of this mode in the disk and the bar, as a function 
of their P.A. with the halo. Applying this procedure to N3, during the first Gyr, 
shows that the disk $A_2$ integrated between 8~kpc -- 12~kpc displays its 
minima/maxima when the disk is oriented at $-45^\circ/+45^\circ$ to the halo, 
i.e., when it leads/trails the DM halo major axis.
During the second Gyr, the $A_2$ has maxima
when the disk is oriented at $-90^\circ$ to the halo in about 1/3 of the rotations. 
In other cases, the
$A_2$ appear to `librate' around P.A.$\approx 90^\circ$. As the disk becomes
less oval due to its interaction with the halo and the bar, the residual $A_2$ always
librates around P.A.$\approx 90^\circ$. 

The same analysis repeated for the bar in N3 shows that early in the evolution, 
the bar `stagnates' at P.A.$\sim 90^\circ$ to the halo. At later times, the bar 
sometimes appears stronger at this angle. But in most cases, $\sim 2/3$,  
$A_2$ appears irregular as a function of P.A. Hence the bar, as the least massive
object in comparison with the disk and the halo, shows a much {\it less} regular 
behavior of its $A_2$. 

We also find that stellar bars forming late in the simulation are shorter, than the
early bars, e.g., in N16, where the bar starts to grow anew after $\sim 6-7$~Gyr
and continues its slow growth until the end of the simulations. In addition, early bars,
large-scale and nuclear, are visibly dominated by the SF and are gas-rich. While 
their late large-scale counterparts appear to be gas-poor.

Fig.~6$b$  exhibits the evolution of the bar
pattern speed in N3. Due to the disk growth from inside-out,  it
shows rather an unusual behavior --- it appears at a fixed, $\sim 90^\circ$, angle
to the halo and remains in this position for a large fraction of a Gyr. Then the
bar speeds up sharply, till $\sim 2$~Gyr.
By that time the central bar-hosting regions in the disk
are mostly built and further addition of the mass to the disk happens well outside.

\subsection{Nested Bars and Spirals}

In the models presented here, the growing 
stellar disks preserve their strong ellipticity
(prolateness) over the first 1--2~Gyr. Strong nuclear gaseous bars form
and experience a distributed SF over their surface. In some models, these 
bars appear tumbling much faster than 
the (oval) disk figure. A pair of grand-design 
spiral arms extend from the surrounding nuclear rings outward, across much 
of the stellar disks. After few rotations, the nuclear gas bar collapses to 
the center (e.g., N18 and N26 models in Figs.~10, 11).
%
\begin{figure*}
\begin{center}
\includegraphics[width=6.0cm,angle=0]{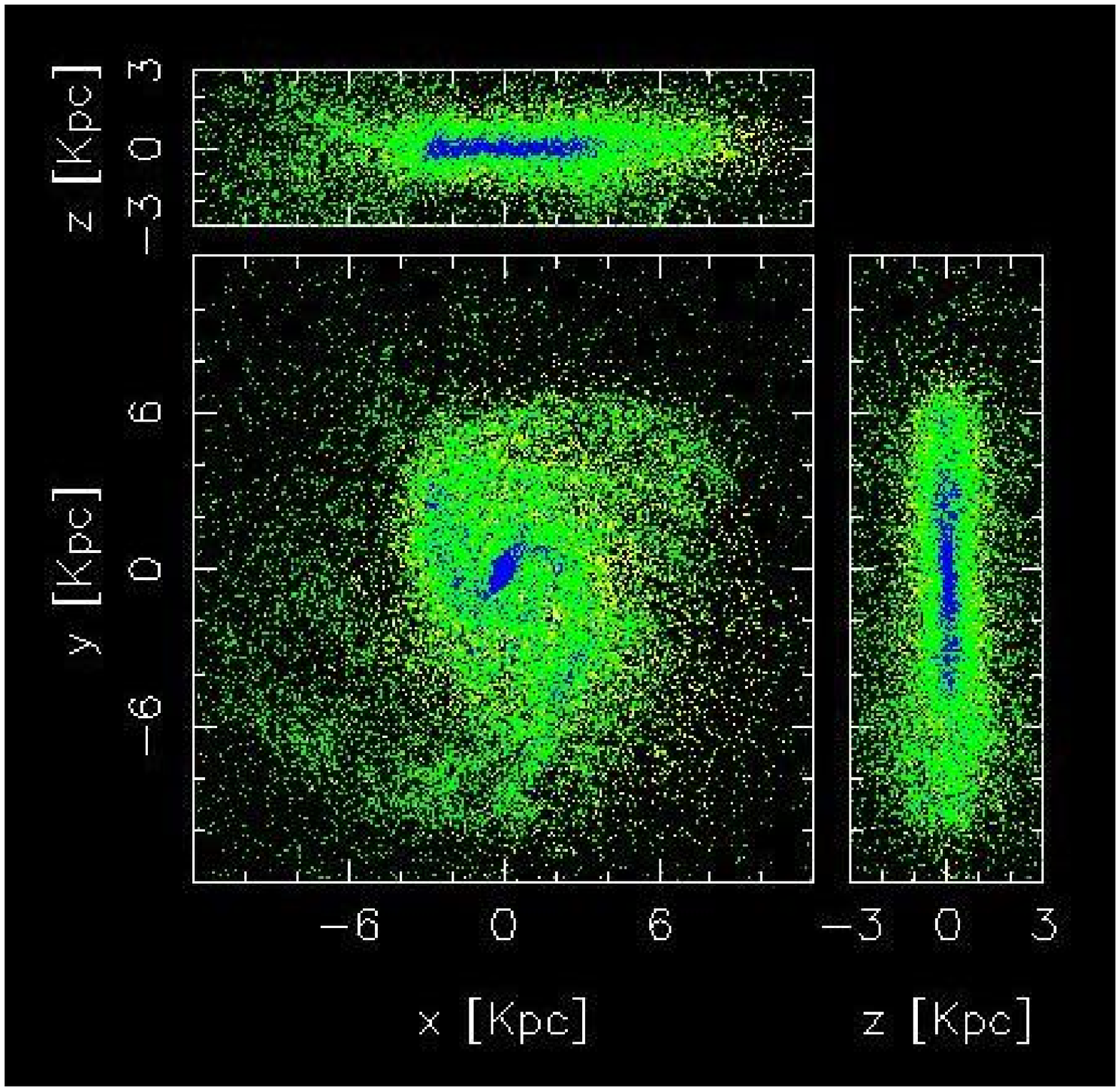}\hspace{0.1cm}
\includegraphics[width=6.0cm,angle=0]{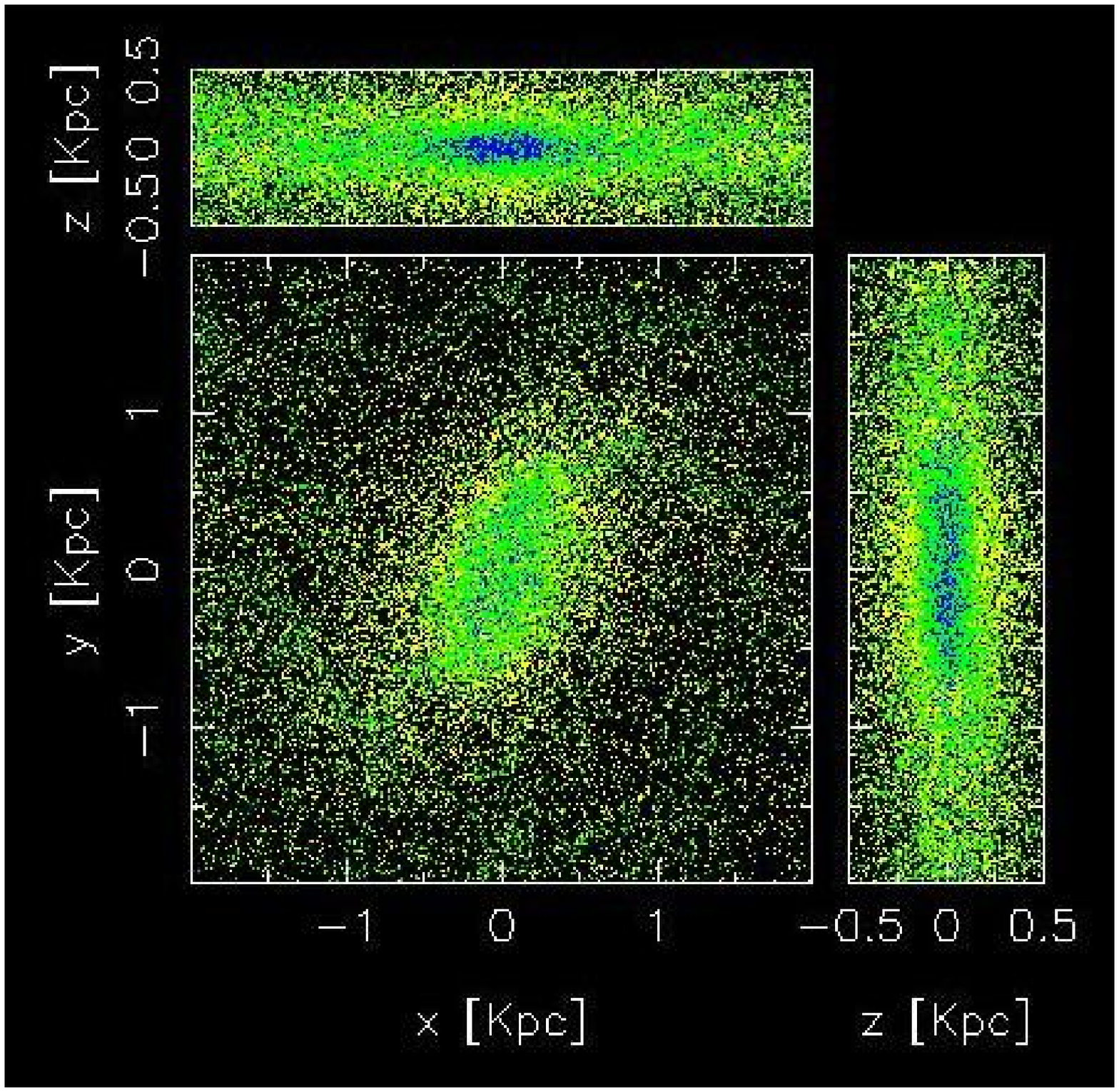}\\
\end{center}
\caption{Nuclear bar tumbling with a pattern speed much higher than the oval 
stellar disk in N18 at $\tau\sim 1.7$~Gyr, seen in gas$+$stars$+$SF 
colors, and shown within the central 12~kpc (left) and 2~kpc (right). 
}
\end{figure*}  

\begin{figure*}
\begin{center}
\includegraphics[width=5.5cm,angle=0]{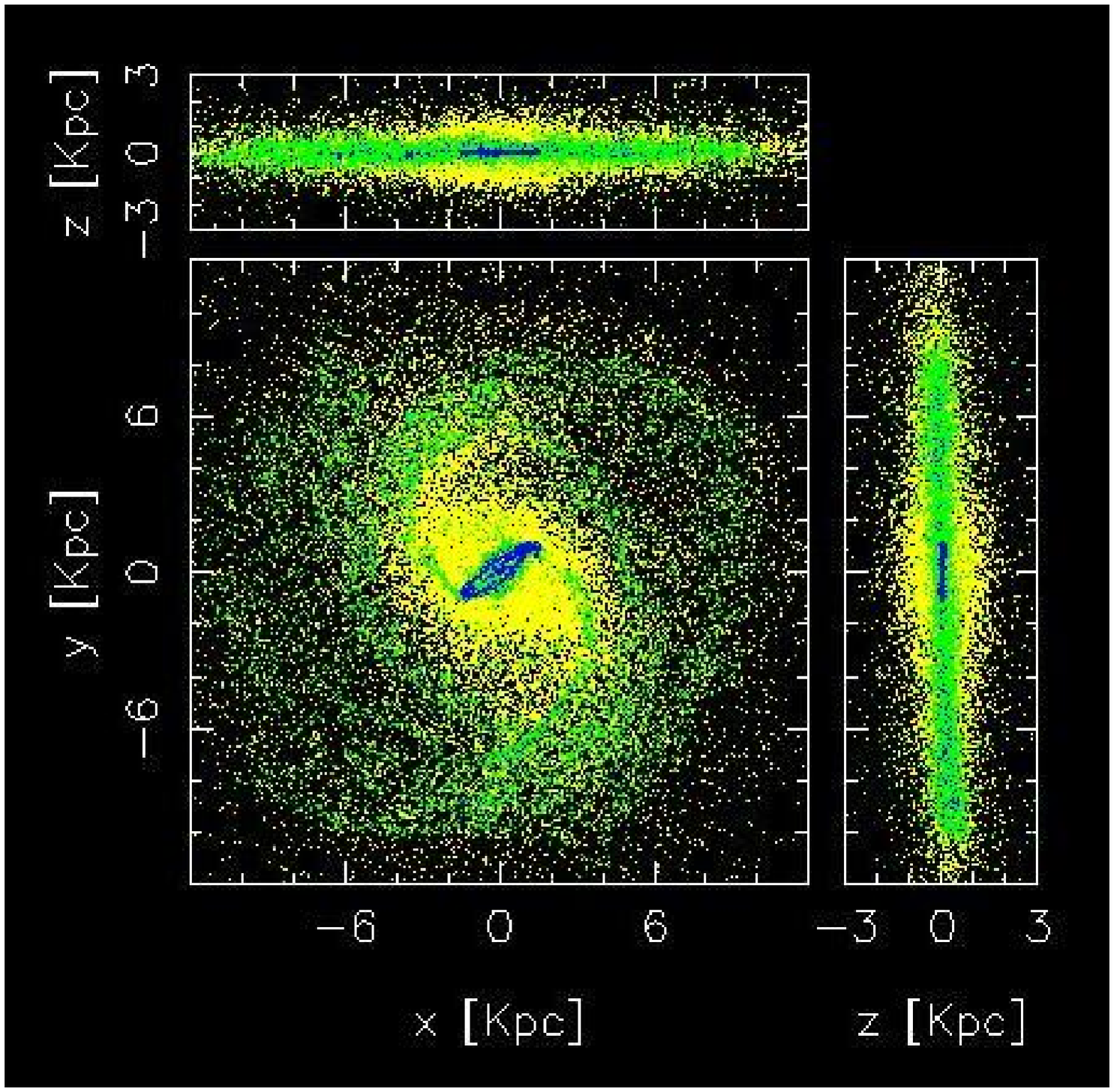}\hspace{0.1cm}
\includegraphics[width=5.5cm,angle=0]{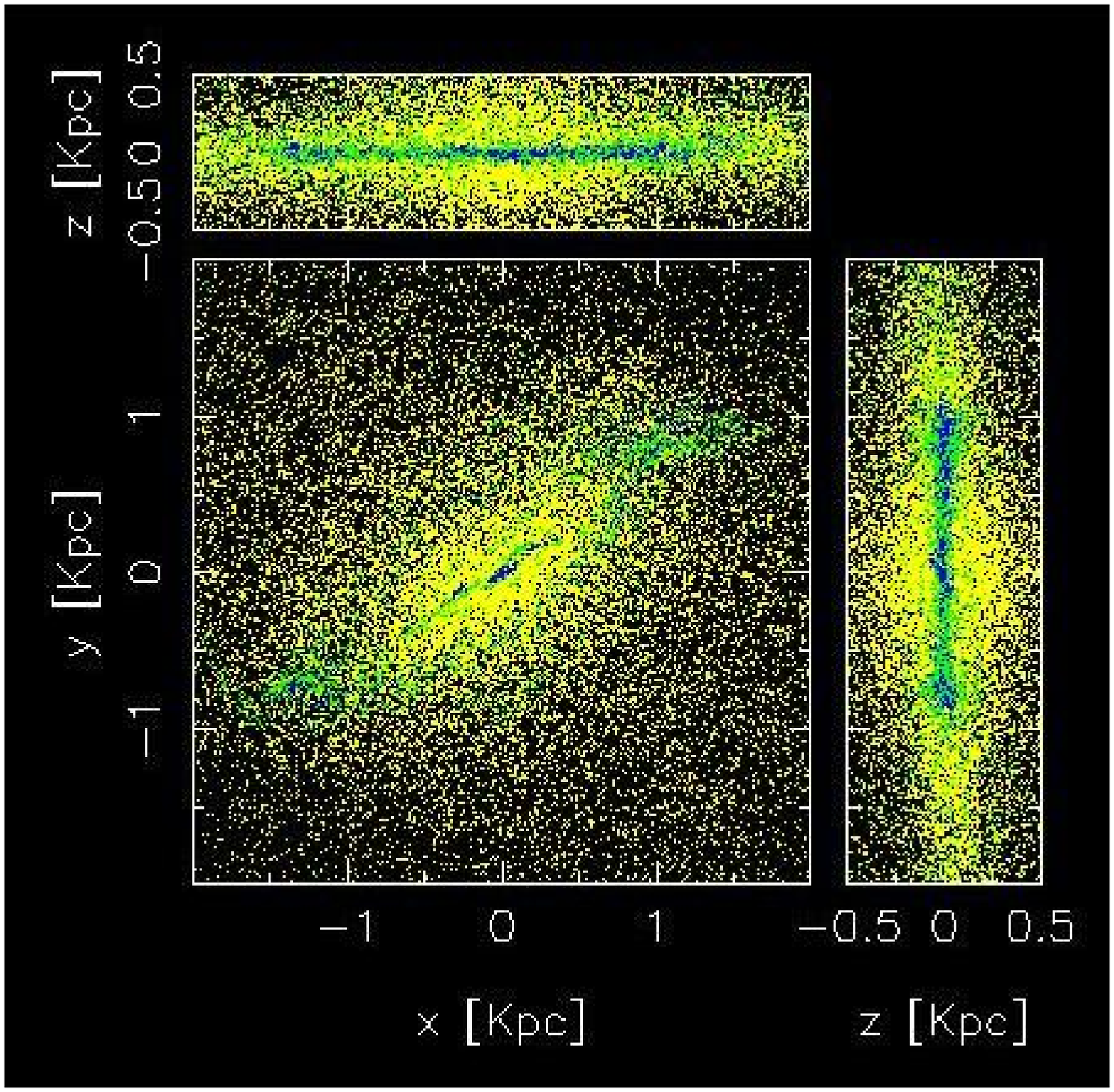}\hspace{0.1cm}
\includegraphics[width=5.5cm,angle=0]{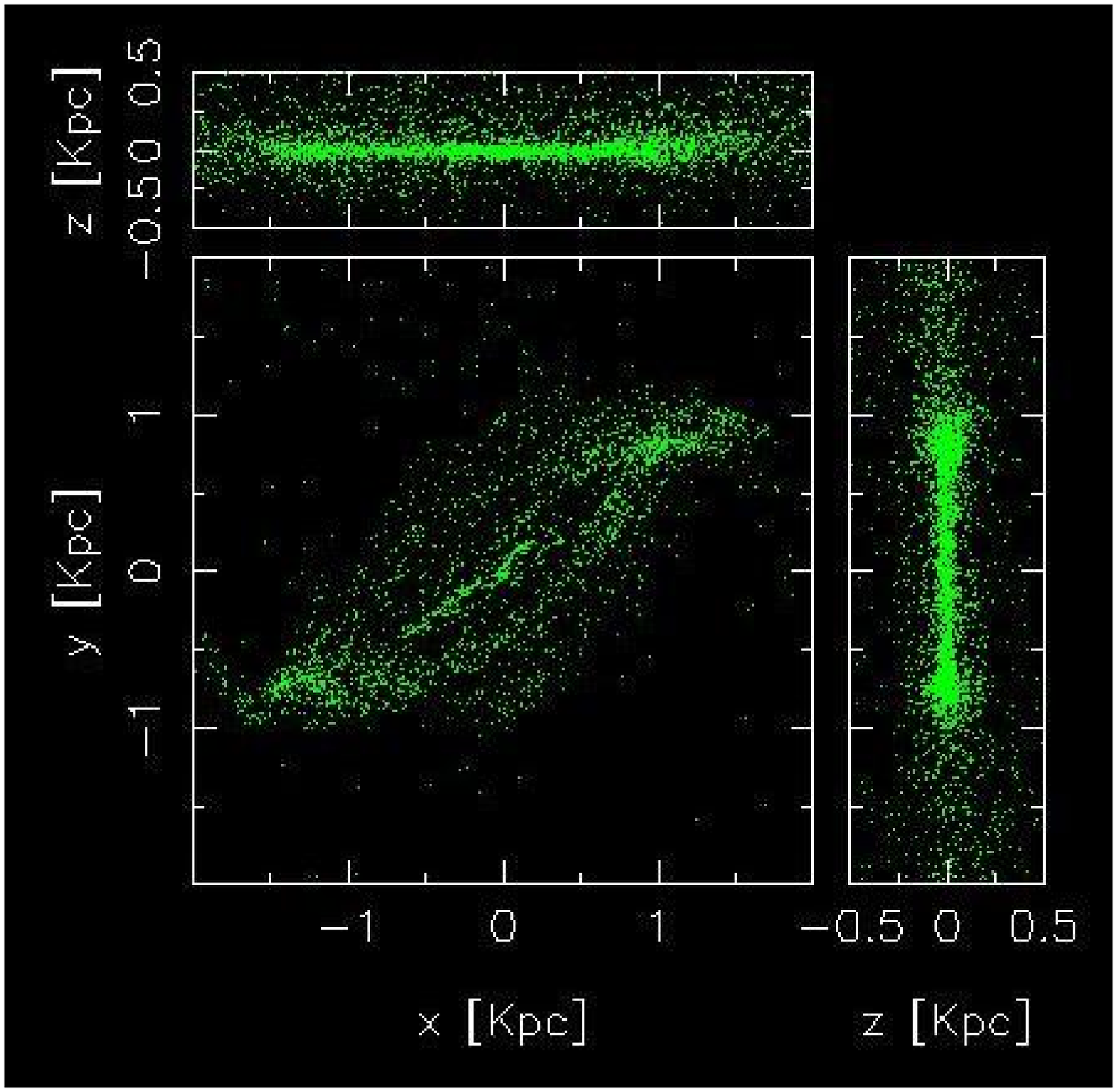}\\
\end{center}
\begin{center}
\includegraphics[width=5.5cm,angle=0]{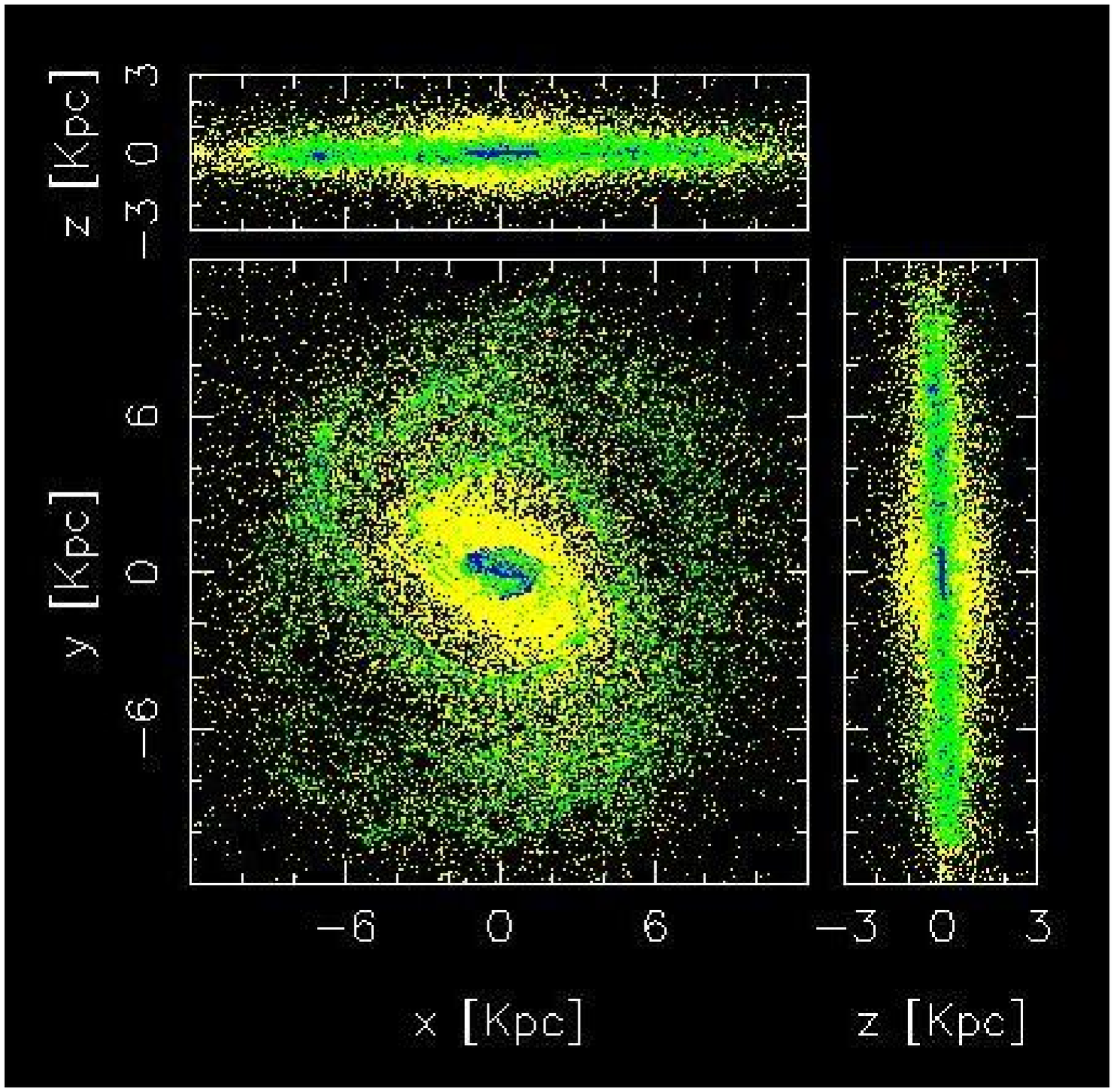}\hspace{0.1cm}
\includegraphics[width=5.5cm,angle=0]{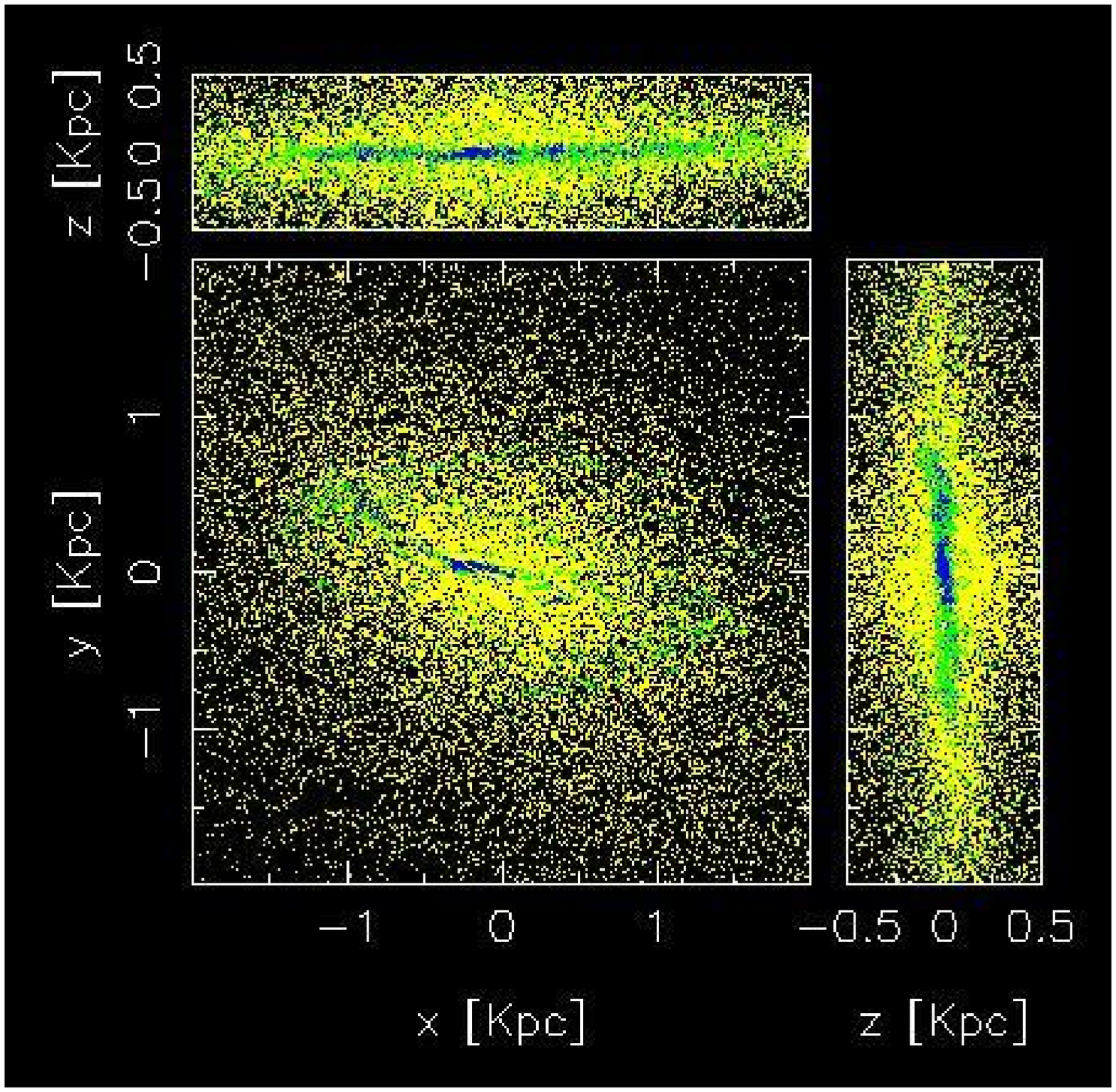}\hspace{0.1cm}
\includegraphics[width=5.5cm,angle=0]{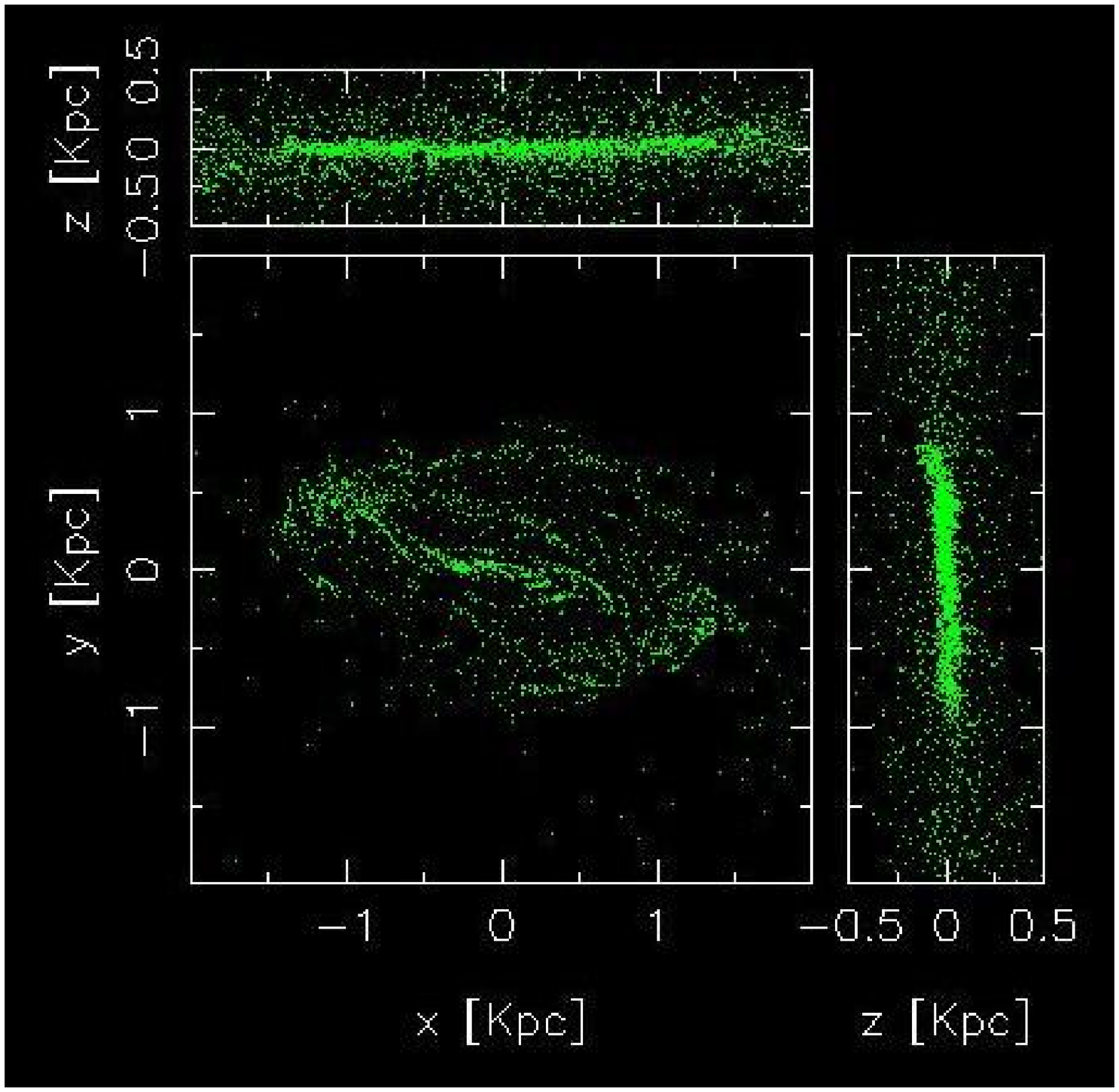}\\
\end{center}
\caption{Differing pattern speeds of nested bars in N26 --- two snapshots. 
\underline{Upper:} large-scale stellar bar in gas$+$stars$+$SF (left),
nuclear bar in gas$+$stars$+$SF (center) and in gas (right) at $\tau\sim 3.00$~Gyr.
\underline{Lower:} same as above but at $\tau\sim 3.07$~Gyr. Left frame sizes 
are $12\times 12$~kpc, center and right frames are $2\times 2$~kpc. Note
that the bars are nearly perpendicular at the former time and nearly aligned
at the latter time.
}
\end{figure*} 
%

These phenomena become progressively more pronounced along the sequence
N16$\rightarrow$N22, excluding N17 and N20. N23--N25 are more clumpy initially 
(corresponding to Shlosman \& Noguchi [1993] scenario with clumps spiraling in 
because of a dynamical friction) and with  no pronounced
nuclear morphology later. The appearance of these clumps can be clearly correlated
with a sharp decrease in the thermalization of stellar feedback energy, down to 
$\epsilon_{\rm SF}=0.05$.

Model N26 represents one of the clearest cases
of nested bars having different pattern speeds (Fig.~11).  
The nuclear bar appears first within the central kpc at
around 1~Gyr as a result of the gas inflow from larger scales, where $m=2$ and 
$m=3$ are clearly visible and constantly evolve. This nuclear bar can be 
observed both in gas (and 
the associated SF) and in stars. At $\tau\sim 1.5$~Gyr, the nuclear gaseous ring 
appears
with an associated SF and is connected to the outer disk by a pair of grand-design
spiral arms. At this time, the nuclear bar seems to have a different pattern speed 
from these arms
and tumbles faster. After $\sim 3$~Gyr, a large-scale stellar bar becomes visible
within the central 5~kpc. The shape and the existence of the nuclear ring depends 
on the mutual alignment of the nested bars (Shlosman \& Heller 2002). 
The ring is also visible in the edge-on 
disk as being dense and much thinner than the gas layer within the central kpc. 
we also confirm that the strength of the small bar depends on the mutual orientation 
of the
bars --- it is stronger when the bars are aligned and weaker when they are normal
to each other (Heller, Shlosman \& Englmaier 2001; Shlosman \& Heller 2002; Englmaier
\& Shlosman 2004; Debattista \& Shen 2006). 

We note that the nuclear bars are gas-rich in the beginning of this process
and form stars vigorously. Gradually the gas in the bar is dumped on the 
center, and the bar dissolves after $\tau\sim 4$~Gyr. The pattern speeds of the
nuclear bars in all models increase over the first $1-3$~Gyr and decline thereafter.
The stellar bar dissolves by 
$\tau\sim 10$~Gyr. Nuclear ring survives and becomes amorphous. A specific case
of evolution of nested bars has been analyzed by Heller et al. (2007).

\subsection{Angular Momentum Redistribution}

The pure DM models conserve the total angular momentum $J$ with the precision of 0.1\%.
In models with baryons, $J$ is not conserved {\it apriori}
because of the processes associated with the SF and feedback. Energy
input from stars leads to pressure gradients, i.e., hydrodynamical torques, etc.
We find that $J$ of all particles in models with SF is
nevertheless a well preserved quantity --- across the models it increases
by a small $\sim 0.3\%-0.4\%$. These processes, therefore, appear much less
important than the gravitational torques for the $J$ balance. 
 
For baryons, $J$ decreases by about 
$25\%-30\%$ over the Hubble time, while for the DM matter $J$ increases by about 
$2.5\%-3\%$. Most of this decrease happens during the first 3--5~Gyr. For stars,
$J$ saturates after $\sim 5-7$~Gyr, while $J$ in the gas continues to 
decrease --- by the end of the simulations about $25\%-50\%$ of baryonic $J$ is in the 
gas, depending on the model. Note that this involves the gas at large radii as well, 
in the regions devoid of stars. This explains the large contribution to $J$ by the 
gas compared to its overall low mass fraction at the end.

\begin{figure}
\begin{center}
\includegraphics[angle=-90,scale=0.4]{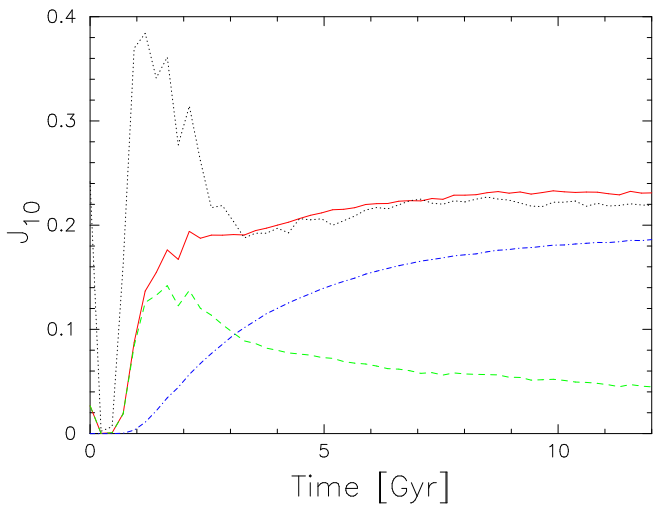}
\end{center}
\caption{Evolution of the angular momentum, $J_{\rm 10}$ within 10~kpc in N3: 
for the DM (black dotted lane),
for the baryons (red solid lane), for the stars (blue dash-dotted) and for the gas 
(green dashed).}
\end{figure}

Of a special interest is the balance of $J$ within the central 10~kpc,
$J_{\rm 10}$, 
where most of the baryonic disk resides. This angular momentum behaves in the
following way (e.g., Fig.~12 for N3). The peak of  $J_{\rm 10}$ in the 
DM is achieved
at $\tau\sim 1.2-1.9$~Gyr with the follow up sharp decrease by a factor of 2--2.5. 
The central 10~kpc remains dominated by $J_{\rm 10}$ contributed by the DM for the first 
$\sim 3$~Gyr. The rise in the baryonic $J_{\rm 10}$ is much slower (than for the DM) 
within this 
region and it reaches the maximum only at $\sim 8$~Gyr, in some models declining 
about 25\% thereafter. The peak of $J_{\rm 10}$ for the DM has a 
similar value, within a factor $\ltorder 2$, to that of the baryons within 10~kpc 
(e.g., Fig.~12). Even more interesting is {\it the close similarity between
the values of $J_{\rm 10}$ for the baryons and DM after the first `splash' over most 
of the evolution time} --- this `equipartition' of $J_{\rm 10}$ tells that the angular 
momentum transfer from the disk to the halo is very efficient and saturates when both 
components have equal $J$. This behavior is observed for all models. 

Stellar $J_{10}$ mimics that of the baryons in total, being about
25\% lower at all times. For the gaseous component, $J_{10}$ reaches
its maximum between 1.5--3~Gyr, within a subsequent sharp decline (by a factor of 2) 
and a slow decline thereafter. The gas $J_{10}$ is always much smaller than 
in the DM (by a factor of 2). The evolution of gas $J$ mimics (somewhat) that of the 
DM overall.  

The specific angular momentum $j$ of the gas is defined as $J_{\rm gas}/M_{\rm gas}$, 
and similarly for stars and the DM. In models with baryons we find that
$j_{\rm DM}$ increases by about $1\%-2\%$ over the Hubble time, most of this increase 
is dated by $z\sim 9-10$. The baryons lose about $15\%-25\%$ of $j$, mainly during the 
epoch $z\sim 4-7$. However, $j$ of the gas
and stellar particles {\it separately} increases (nearly) always monotonically. 
Initially, stars form at the very center with the minimal $j$, then the SF propagates
outward consuming the gas with higher $j$. Again, the specific angular momentum 
$j$ of the gas increases with time in all models because the SF first consumes
particles with lowest $j$. Typically, the stellar $j$ saturates its growth after few 
Gyr, while the gas continues to increase its $j$. Note, that the number of gas 
particles is not decreasing initially, but their mass is --- the algorithm is using the 
option of producing generations of stars from the same gas particles before 
removing the gas from the balance sheet (e.g., Fig.~13). 

\begin{figure}
\begin{center}
\includegraphics[angle=-90,scale=0.4]{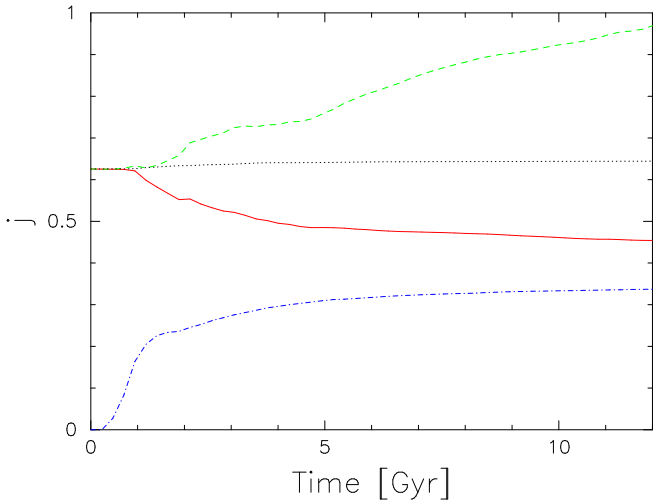}
\end{center}
\caption{Evolution of specific angular momentum $j$ for gas (green dashed lane), stars
(blue dot-dashed), baryons (red solid) and the DM (black dotted) in the N3 model.}
\end{figure}

\section{Results: Comparison of Model Evolution}

Here we describe the model evolution based on the variation of specific
parameters related to the SF processes and to the feedback from stellar
evolution. 

\subsection{Varying the Equation of State: Isothermal Models}

The isothermal EOS has 
been used in N5 and in N22, N24 and N26. Compared to N3, all the isothermal
models develop stronger large-scale ($\sim 3-4$~kpc size) bars and stronger 
nuclear bars. The large bars drive much more pronounced outer spiral arms, both in 
stars and in gas, and accumulate increasingly clumpy gas, especially during the first 
$\sim 1.3$~Gyr. (This is also true when comparing models with and without isothermal
EOS: N22 with N21, and N24 with N23.) These large bars shrink radially 
(arms persist in the gas) to within the central kpc and largely dissolve
due to the quadrupole interaction with the halos, and to a lesser degree
due to the heating by a dynamical friction from the numerous clumps, just to develop 
sometime later on. Compared to N3, N21 and N23, the bars in the isothermal models show 
a higher rate of SF. The nuclear bars persist for a few Gyrs, leaving an oval 
distortion thereafter. Similarly to the large-scale bars, nuclear bars are `resurrected' from 
time to time. 

In all of the above models a central opening of about 1~kpc in the gas 
appears, surrounded by a ring, by $\tau\sim 5$~Gyr. Gaseous
bars driving two gaseous spirals develop within this hole, from time to time, and 
exhibit a faster pattern speed than the surrounding flocculent spiral structure. 
By $\tau\sim 7.5$~Gyr, the hole grows to about 2--3~kpc in radius and more prominent in 
the models with the isothermal EOS. The comparison
of pattern speeds shows a diverging evolution for the gas inside 1~kpc and 5~kpc,
e.g., late (after 9~Gyr) speedup of the gaseous component in N26, or a late gaseous
bar in N22. 

Overall, the isothermal models show morphological features that are 
sharper, including nuclear rings and bars, and their 
SF rates exceed that of the non-isothermal. The spiral arms outside the 
nuclear rings have a grand-design appearance. During the maxima of the SF, the 
gas disk is more puffed up.  
The edge-on gas disks exhibit a characteristic concave shape during the first 3~Gyrs, 
and the SF is more pronounced in the mid-plane. In comparison, the N3 gas disk is 
more convex. The outer massive spirals in the isothermal models appear to be responsible 
for these differences, which disappers by $\sim 3$~Gyr. The bulges in the
isothermal models extends vertically beyond the disk
after $\tau\sim 2.3$~Gyr, and, by the end of the simulations,
are somewhat more pronounced. The disks in non-isothermal models are smaller,
e.g., in N3 by about 20\%, both in stars and gas.
Typically, much more gas remains outside 
the N3 disk, both in the mid-plane and above the disk. The disk shapes differ 
dramatically during the evolution between non-isothermal and isothermal models and 
the morphology of spiral arms looks different as well.  

\subsection{Varying Feedback From Stellar Evolution, $\epsilon_{\rm SF}$}

In order to investigate the effect of a stellar feedback onto the gas, we have 
varied $\epsilon_{\rm SF}$ from 0.3 in N3, to 0.1 in N17. Additional relevant 
sequences in Table~1 are the `doublets' N8 ($\epsilon_{\rm SF}=0.1$)$\rightarrow$ 
N6 (0.05), N14 (0.3) $\rightarrow$ N15 (0.1), and the isothermal N22 (0.1) 
$\rightarrow$ N24 (0.01), and the `triplet' N21 (0.1) $\rightarrow$ N25 (0.05) 
$\rightarrow$ N23 (0.01). This parameter has a major impact on the evolving 
morphology in our models. The SF starts earlier and tends to be more patchy, 
underlining the clumpy character of the gas, when $\epsilon_{\rm SF}$ is
smaller. The clumps spiral in toward the center because of the dynamical friction,
especially around $\tau\sim 1-2$~Gyr. Models with a smaller feedback also show 
a more prominent morphological features, like bars, nuclear rings and spiral arms. 

In the first doublet sequence, the N17 evolution is similar to that in N3, except 
the SF extends to larger radii. The N3 disk is the smallest among the models 
listed above and the N23 disk is the largest. In the early stages, the central gas 
concentration and its central surface density are observed to increase with decreasing 
$\epsilon_{\rm SF}$. The gas layer thins along the same sequence. 

\begin{figure*}
\begin{center}
\includegraphics[width=5.5cm,angle=0]{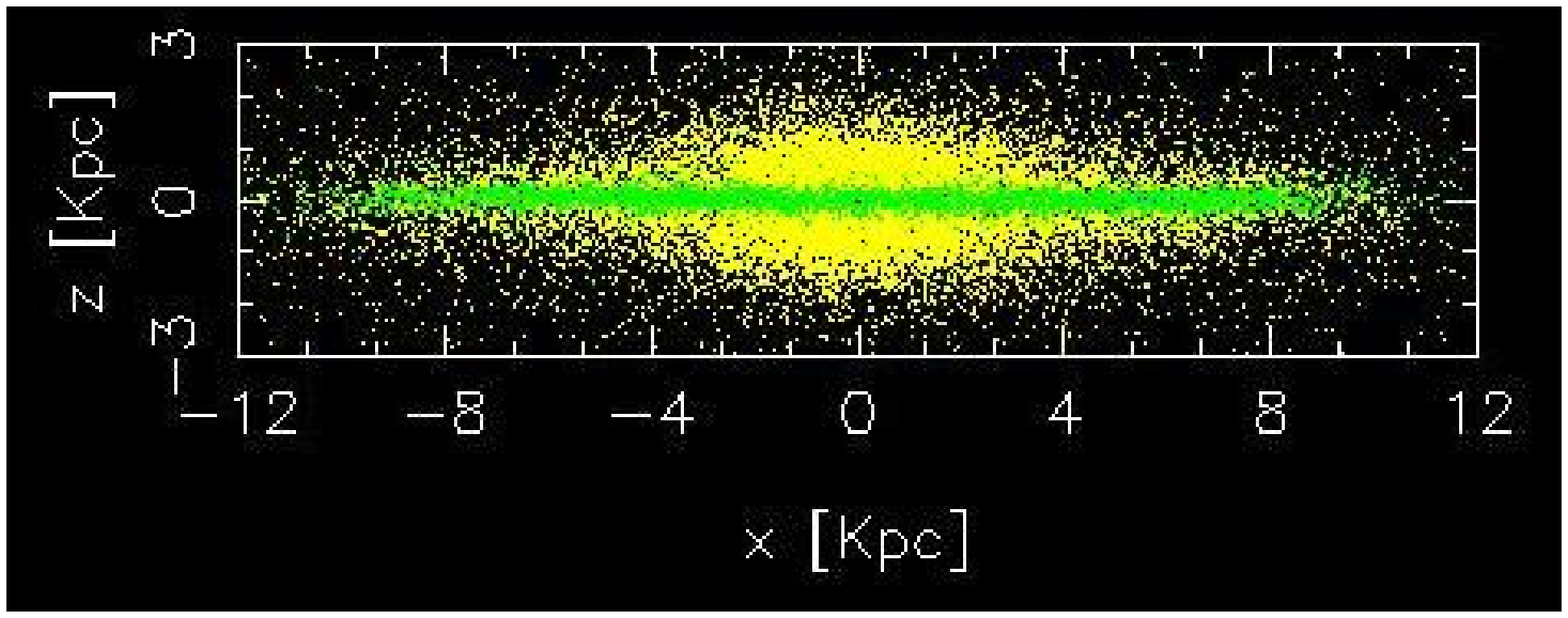}\hspace{0.1cm}
\includegraphics[width=5.5cm,angle=0]{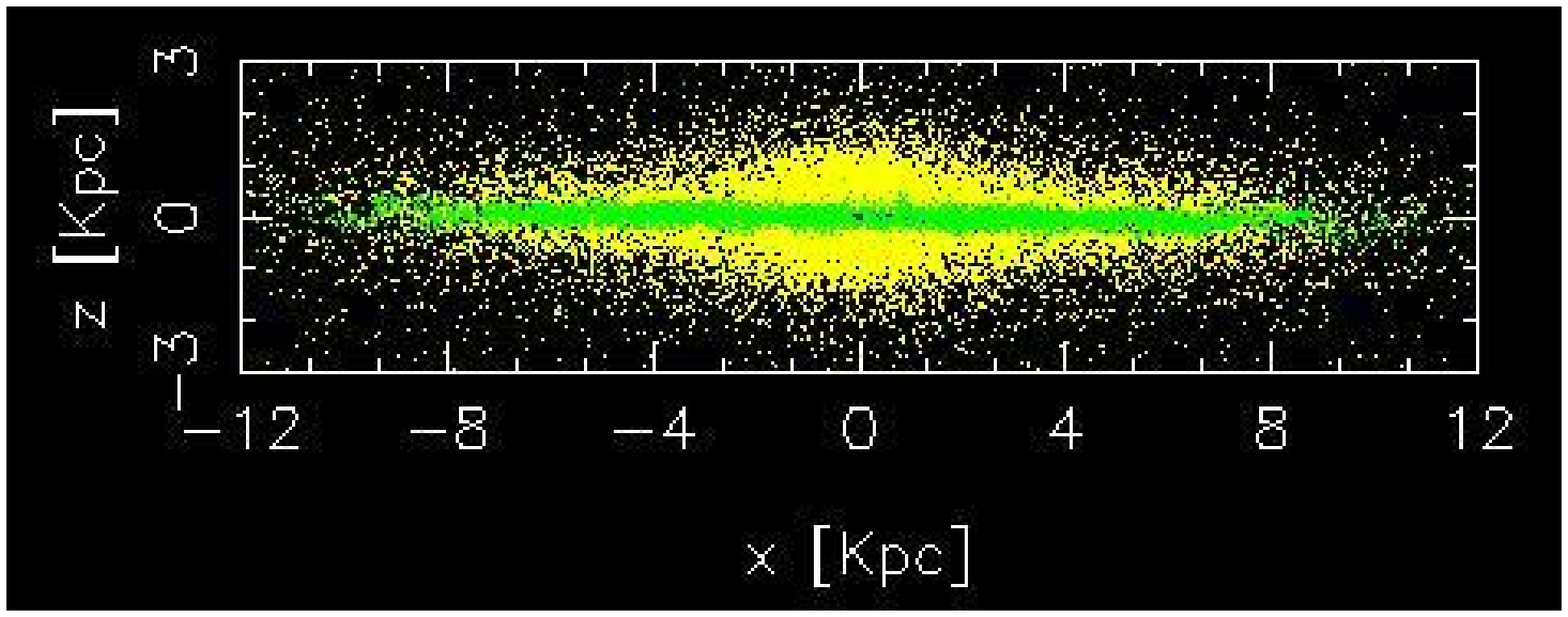}\hspace{0.1cm}
\includegraphics[width=5.5cm,angle=0]{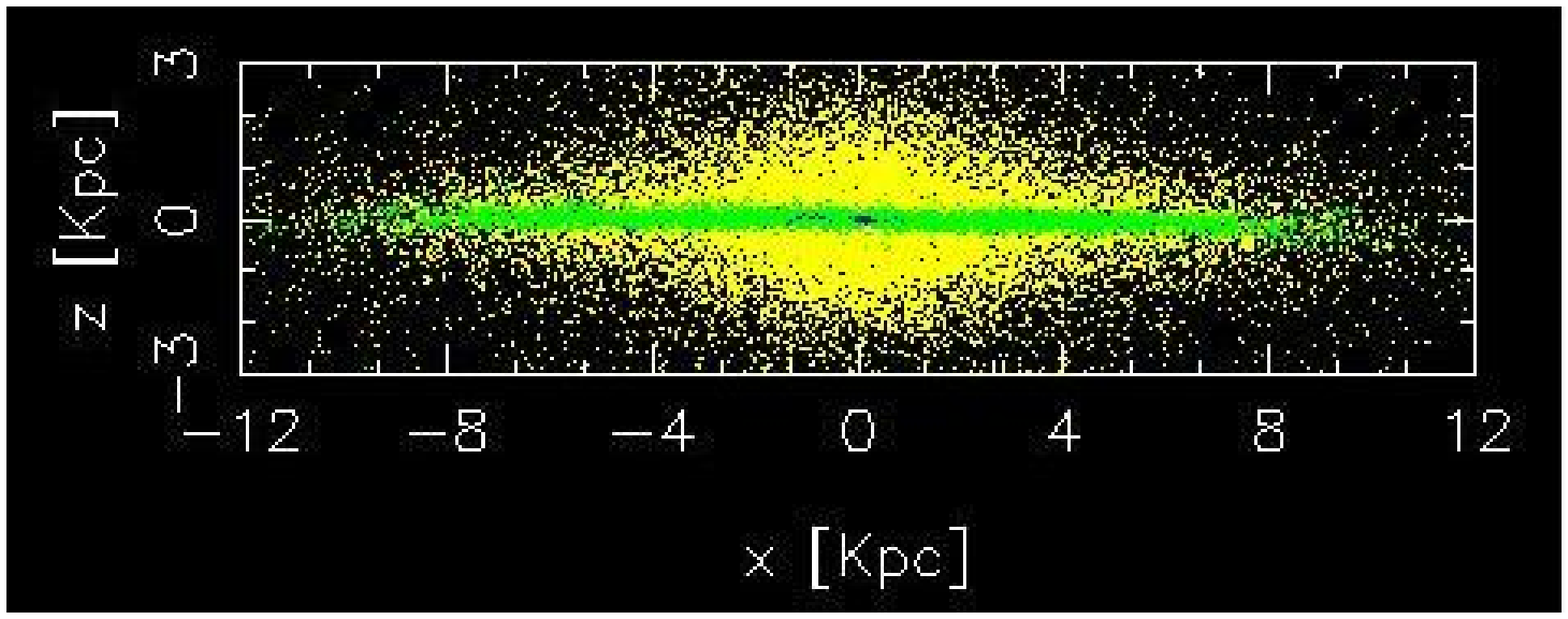}\\
\end{center}
\caption{Stellar bulges in the edge-on disks at the end of the simulations
for models N21 (left), N25 (center) and N23 (right). This sequence 
corresponds to a decreasing value of the feedback parameter, $\epsilon_{\rm SF}$ 
(see Table~1). 
}
\end{figure*} 

The stellar bulges form somewhat earlier in models with lower $\epsilon_{\rm SF}$, in
N23 by $\tau\sim 1.6$~Gyr, in N25 by $\sim 1.9$~Gyr, and in N21 by $\sim 2.5$~Gyr. 
Bulges in N17 and N3 form by $\sim 2.5$~Gyr. By the end of the simulations,
the largest bulge is found in models with the smallest $\epsilon_{\rm SF}$,
among the triplet and doublet parameter sequences given above, i.e., in N23 
(Fig.~14) and N24. The smallest bulges are found in N3, N22. Such a bulge is also 
expected in N21. 
However, as can be seen in Fig.~14, the bulge of N25 is smaller than that of N21.
The reason for this behavior lies in that N21 develops a strong late bar, while 
N25 exhibits only an oval distortion after the first 5~Gyr. N23 lacks a late bar 
totally. Stellar bars are known to heat up the central disk regions and inflate
them vertically by increasing their velocity dispersions, both dynamically and 
secularly (e.g., Athanassoula \& Misiriotis 2002; Berentzen et al. 2007).  
Additional heating exerted by the bar in N21 contributes to the appearance
of a more prominent bulge there. 

We infer that two processes, at least, emerge as being responsible for the bulge 
and stellar bar evolution in our simulations --- both of these processes are
`guided' by $\epsilon_{\rm SF}$, the stellar evolution feedback onto the gas.
We observe that a decrease in $\epsilon_{\rm SF}$ alleviates the clumpiness in the gas,
which in turn heats up the stellar component in the disk, therefore, delaying
or damping the bar instability there. This explains why N21 bulge does not `fit'
in the sequence shown in Fig.~14.
We conclude, therefore, that by the end of the simulations
there is a clear anti-correlation between the bulge prominence and 
$\epsilon_{\rm SF}$.

\subsection{Varying Critical Density for Star Formation, $\alpha_{\rm crit}$}

We look for the effect of the critical gas density for the SF by
comparing N14 ($\alpha_{\rm crit} = 0.5$) with N27 (0.05), and N15 (0.5) with N16
(0.25). The main difference observed along this sequence is the delayed
evolution in N27, where $\sim 3$~kpc bar develops
much later compared to N3 and N14, after $\sim 8$~Gyr only. It can be also 
seen in the SF colors.  The dominant SF in the central 
kpc of this model also appears very late, after $\sim 2$~Gyr, and spreads 
out in $r$ and in $z$ (low surface and volume densities in the gas). 
It picks up, however, after $\sim 5$~Gyr, and remains 
much higher than in other models.  The stellar disk is clearly much hotter 
and puffy in N27. The bulge develops late but becomes more pronounced than in N14.

\begin{figure*}
\begin{center}
\includegraphics[width=5.5cm,angle=0]{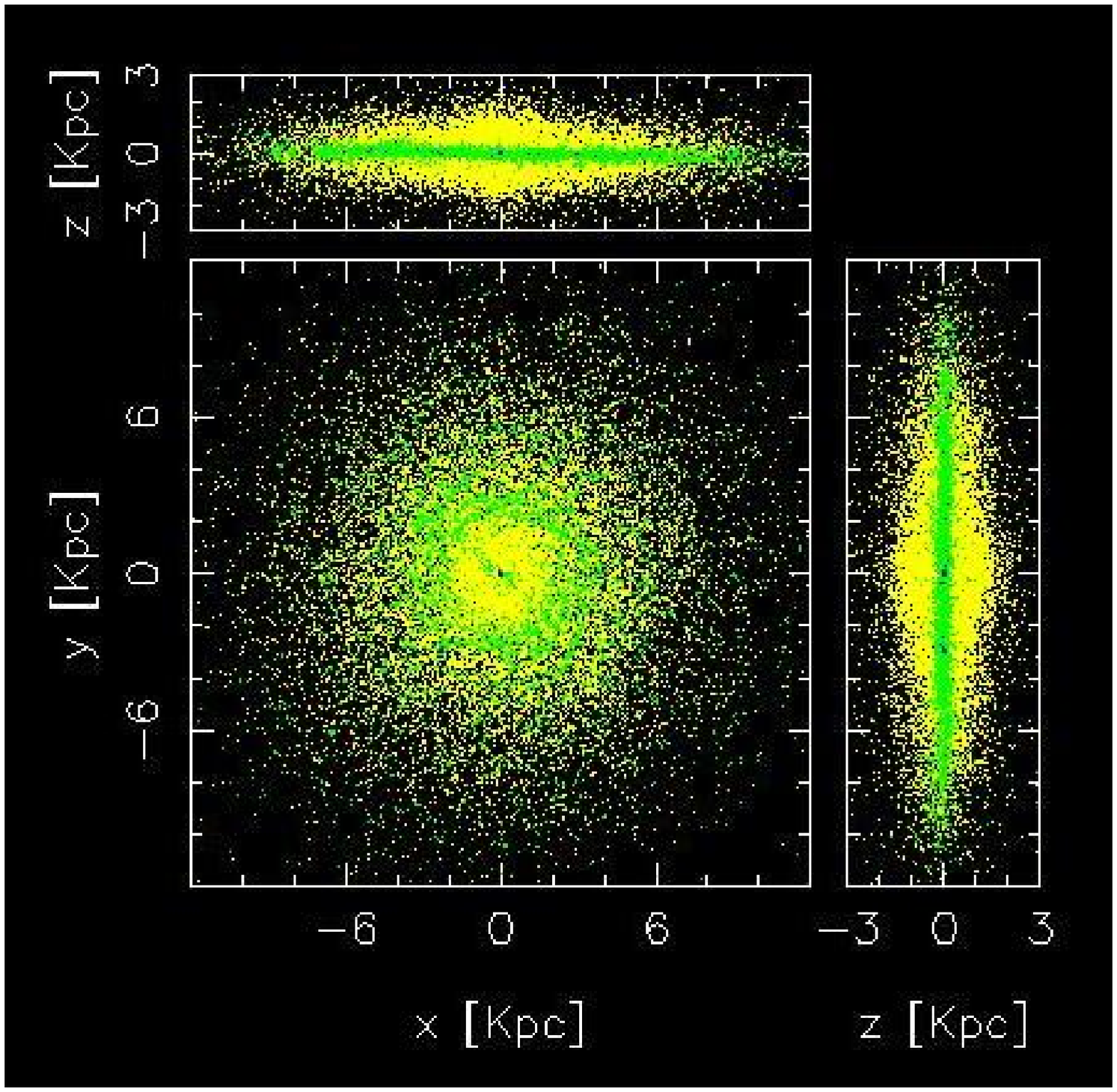}\hspace{0.1cm}
\includegraphics[width=5.5cm,angle=0]{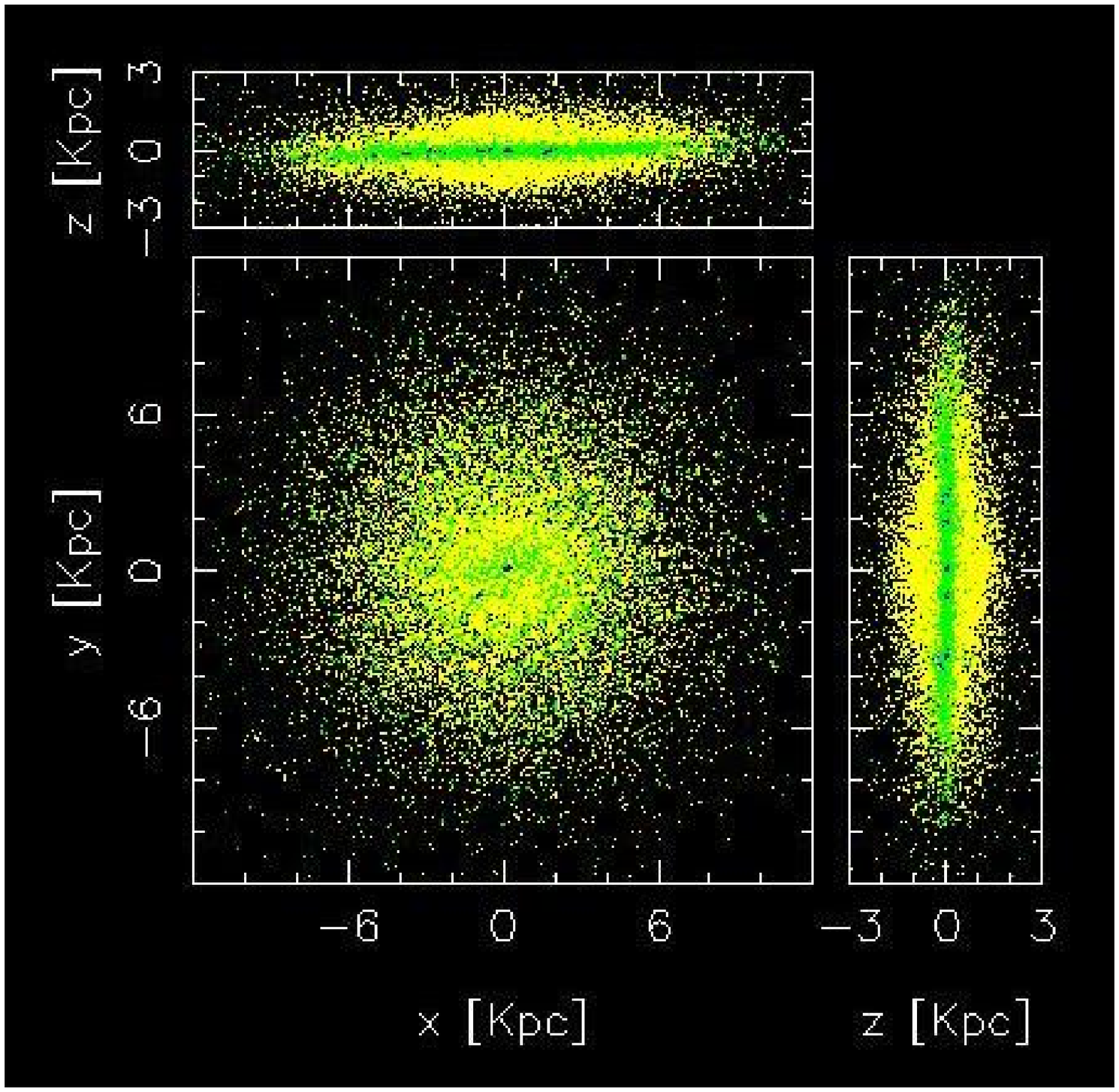}\hspace{0.1cm}
\includegraphics[width=5.5cm,angle=0]{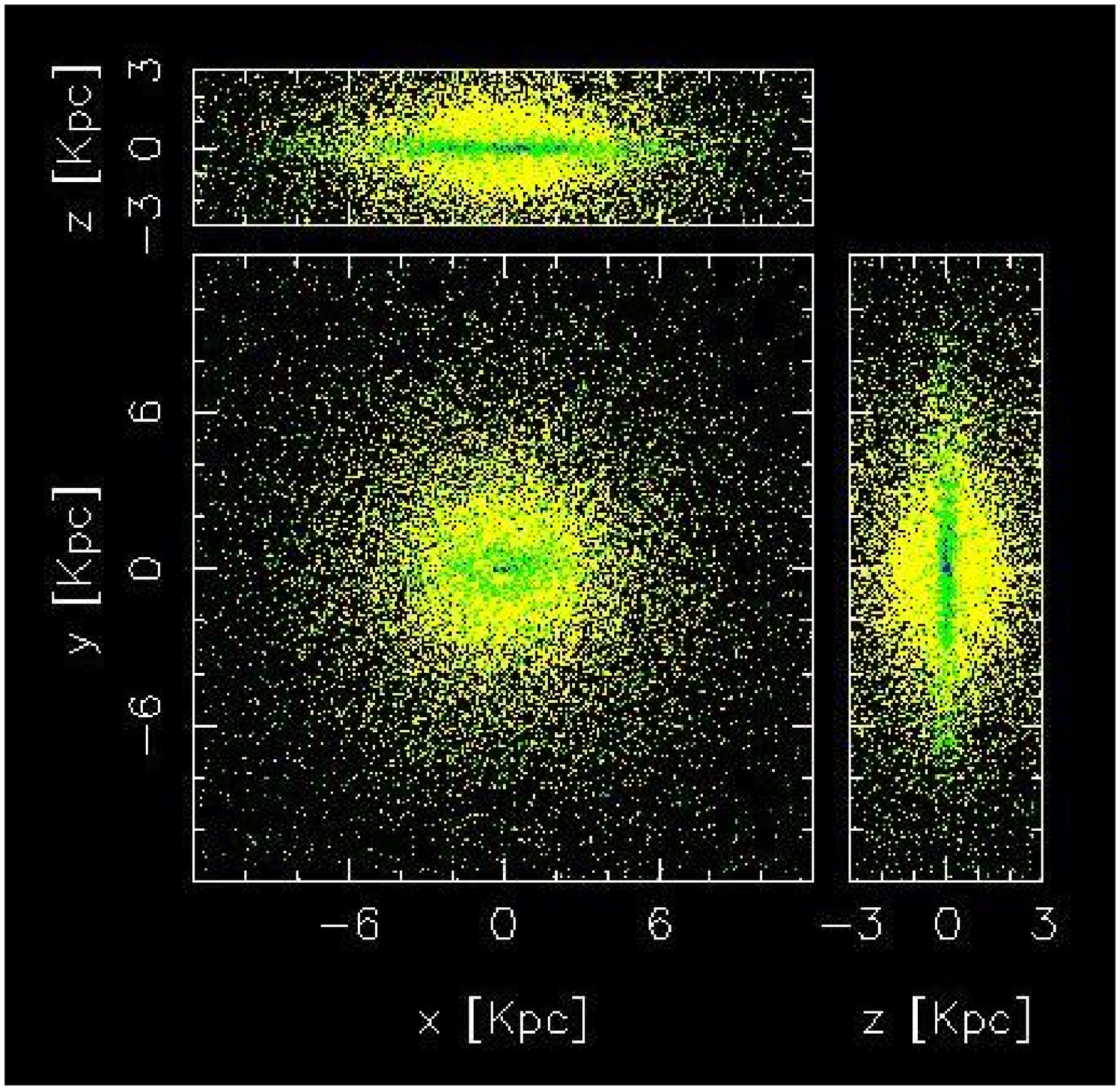}\\
\end{center}
\caption{End disks in N3, N14 and N27. The colors are: stars (yellow), gas (green) 
and SF regions (blue). The N14/N27 sequence of models corresponds to a decreasing
critical density parameter, $\alpha_{\rm crit}$ (see Table~1).
The N3/N14 sequence corresponds to a decreasing $\alpha_{\rm ff}$ parameter.
}
\end{figure*} 

On the other hand, using a low value $\alpha_{\rm crit} = 0.05$ in N27 produces the
smallest $r\sim 4-5$~kpc disk with a very high surface density, enveloped
in a low density stellar and gas material (Fig.~15). The SF is concentrated in
the disk mid-plane. A bar-like stellar and gas response to the halo prolateness
is evident initially within the central $\sim 3$~kpc, but decays after 3~Gyr.
This response forms at a fixed, nearly normal angle with the DM major axis. 

Hence, using a low density threshold for SF, in the presence of a substantial feedback,
heats up the stellar and gas components in the growing disk. N27 
is probably the only model where the SF is distributed and is triggered at 
all radii $<15$~kpc in the disk --- this disk does {\it not} appear to grow 
from inside out. It produces a substantial bulge but the smallest stellar disk. 

\subsection{Varying Timescale for Star Formation, $\alpha_{\rm ff}$}

A number of models have a much shorter timescale for SF, 
$\alpha_{\rm ff} = 1$ instead of 10 (see Table~1) --- we compare N3
($\alpha_{\rm ff} = 10$) with N14 $\alpha_{\rm ff}=1$, N8 (10) with N16
(1), and N17 (10) with N15 (1). 

Early enough, by $\tau\sim 0.9$~Gyr, the SF in N14 is less concentrated toward
the midplane and the stellar and gas disks are visibly thicker than in N3. The 
SF is also more profound in the N14 bars. The gas
surrounding the disk in N14 is more noticeable, to the extent that the gas disk
flares beyond 5~kpc. By $\sim 1.4$~Gyr,
a 3~kpc disk of higher surface density emerges in N14 --- while the vertical
thickness of the disk is now comparable to N3, the radial size of N14 disk
is smaller. By $\sim 7$~Gyr the bulge of N3 seems more pronounced, while toward
the end the N14 bar is more gas-rich than in N3.

N15 shows earlier and higher rate of SF and has an initial stellar, gas (and SF) 
bar. This bar weakens and strengthens occasionally but after $\sim 5$~Gyr grows 
monotonically. For a comparison, N16 develops such a large-scale bar by 7~Gyr.
The disk in N15 is more puffed up than in N17 and the grand-design
spiral arms are much less visible.  
Grand-design spiral structure is strong between $\sim 2.5-5$~Gyr --- apparently
when the bar/disk are aligned at certain angles to the DM halo major axis. The
SF, which delineates the arms, follows them, while it is more diffuse in N17.
The disk is thick and concave at this time, much more than in N17. Compact gas clumps 
are present thereafter. By 3.5~Gyr, the SF in N15 has subsided more than in N17, the
disk seems to be somewhat smaller and the bulge is more compact as well. It also
has less gas in the central kpc. The gas layer starts to thin first in this model
compared to N17. Towards the end of the simulations it is clear that the disk in
N15 is overall thicker, and the bulge is standing out with respect to the face-on 
disk more than in N17. This bulge has an oval/barlike shape. 

Overall, models with smaller $\alpha_{\rm ff}$ form somewhat smaller in size but
more puffed up stellar disks.
 
\subsection{Standard Model with an Imposed Axisymmetric Halo}

To enforce the halo axial symmetry during a continuous infall of material, 
we have randomized the azimuths of 5\% DM particles, every timestep. 
Therefore, no initial bars are either triggered or driven by the halo prolateness.
Furthermore, this should lower the ability of the halo particles to be in 
resonance with the nonaxisymmetric features in the disk. This, in turn, 
should decrease the efficiency of angular momentum flow from the disk to the 
halo and change the stellar bar evolution. 

Indeed, in model N4, the initial collapse proceeds without the `Cat's Cradle'
phase and no DM filaments are present. Consequently, the gas does not form the
massive clumps. The disk remains nearly axisymmetric at all times and no
substantial spiral structure is visible in the outer part, unlike in in
all other models. Some stronger 
oval distortion develops inside 1--2~kpc after $\sim 8$~Gyr for some time and 
decays thereafter. The gas bars are visible from time to time in this region.
This evolution is in a sharp contrast with N3 which
displays an early and late bar activity on various spatial scales. The SF and
spiral morphology in N4 also differ substantialy.

\subsection{Varying Gravitational Softening, $\epsilon_{\rm grav}$}

\begin{figure}
\begin{center}
\includegraphics[width=6.5cm,angle=-90]{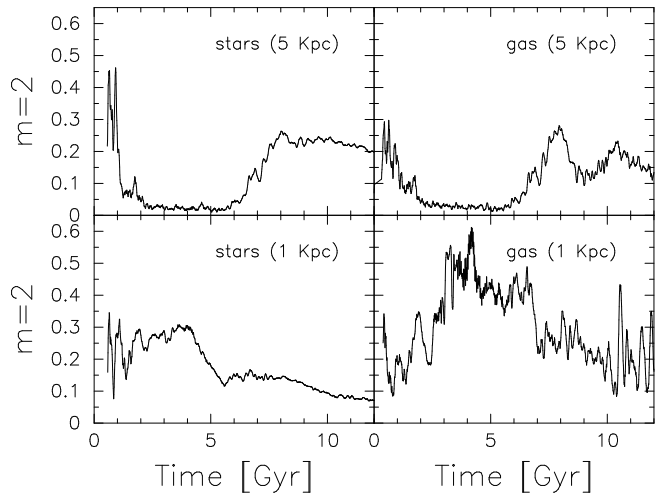}\\
\end{center}
\caption{The evolution of $m=2$ Fourier component $A_2$ in the N18 disk. Shown are 
stars (left frames) and gas (right frames) within the central 5~kpc (upper frames) 
and 1~kpc (lower).}
\end{figure}

Models N18 and N19 have larger gravitational softening in gas and in stars ---
$\epsilon_{\rm grav}=250$~pc instead of 150~pc in other models. We first compare 
N18 with N3.  
The general evolution of N3 and N18 is similar but it differs in some details. 
First, the former model shows sharper features (shocks, spiral arms, SF regions). 
Second, the
gas there is somewhat more clumpy and is able to concentrate more towards the
center and the disk mid-plane. Already during the first Gyr, N3 gas shows a larger
surface density in the central kpc. Third, the bars that form in this model are less 
strong than in N18 (Figs.~6, 16). While the N3 nuclear bar is gas poor (efficiently 
channeling it to
the center), the N18 bar remains gas-rich for a few Gyrs. Two factors may reduce
the bar strength here: larger gas and stellar concentration to the center and
more clumpy gas that scatters bar stars. This latter feature can also prevent
the bar orbits to lock in the resonance with the DM halo orbits, thus resulting
in a weaker bar. Finally, the stellar bulge in N3 is less prominent than in N18 but 
is more centrally concentrated.   

Next, we compare N21 with N19. The stellar softening is the same but the gas
softening is 150~pc in N21 instead of 250~pc in N19. As in the previous pair,
the morphological features are more detailed in N21. This is noticeable 
in the distribution of SF and gas in the strong spiral arms --- both are
more clumpy, but not in stars. As before, N21 develops a more massive gas concentration
in the center, but the difference between the models is not as obvious, apparently
because the stellar components have identical softening. The nuclear rings
appear (vertically) thinner in N21 and are much more long-lived. Occasionally,
they collapse to the center. In the long run, the 
difference emerges also in the stellar distribution
as well --- the bulge is visibly larger in N19 and less centrally concentrated.
The bars in N19 are also more gas rich, but periodically this property is inverted
between the models. On the other hand, the bars in N19 appear stronger.
Hence, the gravitational softening emerges as an important parameter which
drives the evolution in our models. Especially in gas, this parameter reflects
the overall smoothness of the gas and can be related to a number of intrinsic
processes in the ISM, heating/cooling and the topology of the magnetic fields ---
those affect the gas clumpiness.
 
\section{Discussion and Conclusions}

We have simulated the formation of a two-component galactic disk within an
assembling DM halo, including the accompanying SF and the stellar feedback
onto the ISM. For this purpose, we have followed the collapse
of an isolated cosmological density perturbation with the spin parameter 
$\lambda=0.05$, from its linear regime at $z=36$ and up to $z=0$. Our 
approach has several advantages. We have improved the SF 
criterion, accounting not only for the contraction and the critical density
of the SF region, but also for the background density. Each gaseous particle
can make up to four generations of stars, with different metallicities,
while an additional parameter allows to consider a delayed collapse of the
molecular clouds due to MHD turbulence. We consider the energy feedback both from
stellar winds and from SN, and we are particularly careful in the way this
energy is deposited in the gas.

We have employed a reasonably high number of particles, allowing us to resolve
a number of important features. In particular, our simulations are fully
self-consistent, in order to avoid artifacts that could be introduced, e.g., by a
rigid halo (e.g., Athanassoula 2002). This enabled us to study interesting
dynamics linked to the halo shape. We have voluntarily refrained from using
a yet higher number of particles, in order to be able to run a large number
of simulations, much more than is usual in similar studies. With our $\sim 30$
models, we tried to gain insight on how the various parameters, and particularly
the SF and the energy feedback, affect the structural properties of the bulge,
the disk and its components, like bars and spirals. Of course, the available
parameter space is very large and much further work and many more simulations
are necessary before any thorough understanding is reached. Yet, even with
the number of simulations we have at our disposal, several clear trends emerge.

In all models the disk forms over a period of time of $2-3$~Gyr, i.e., it reaches
50\% of its final mass over this time, although its
equatorial plane is established very early. The DM halo develops a characteristic
triaxial shape, i.e., prolate in the plane perpendicular to the original
angular momentum, and flattened along the rotation axis. For pure DM models,
the axial ratios of the inner ($< 20$~kpc) halos lie between $b/a\sim 0.8-0.9$ and 
$c/a\sim 0.75-0.85$, and of the outer ($> 50$~kpc) halos between 
$b/a\sim 0.6-0.7$ and $c/a\sim 0.45-0.55$. At the same
time, we find that the halo figure tumbles very slowly, $\sim\pi$ over the
Hubble time. {\it The DM collapse pumps most of the angular momentum into the
internal circulation within the halo, while the baryonic collapse is stopped
by the centrifugal barrier in a flat disk.} Thus the baryons, due to dissipation,
collapse much more profoundly. 

The halo triaxiality decreases with time during the disk growth. Two processes
contribute to this --- (1) the increasing central mass concentration in the
model and the appearance of of numerous scattering centers (blobs) in the halo,
and (2) the out-of phase response of the baryons in the disk to the halo
shape. The first effect reduces somewhat both the flatness and prolateness of 
the halo, even in the absence of a baryonic component, much more so in the presence 
of baryons. The second effect results from the
negligible tumbling of the halo figure --- under these circumstances the inner ILR
(if it exists) and the outer ILR are pushed to the center and to large radii, 
respectively, and the disk acquires
an elongation perpendicular to the major axis of the halo, thus diluting its
potential in the equatorial plane. This effect is model independent and is
unrelated to dissipation.

The angular momentum ($J$) flow within the evolving models has been always directed
outwards ---
the inner disk is losing $J$ to the outer disk and to the surrounding DM halo. 
Overall, the baryons lose their $J$ while the DM increases it in the process of
evolution. On the average, the DM halo acquires about 2.5\%--3\% of $J$ while the
baryons lose about 25\%--30\% of their original momentum. The total $J$ is conserved
to less than 0.5\% even in the models with SF.
The specific angular momentum ($j$) behaves in a more complicated
manner --- for the DM it is almost constant over the Hubble time. For baryons,
$j$ decreases with time but each of the components separately, the gas and stars,
increase their $j$ substantially over the time. This happens because the first
stars form in the center with a minimal $j$, and the first gas consumed by the
SF has small $j$ as well. 

The disk evolution can be roughly divided into two stages: in the early stage,
the baryons flow towards the forming small and amorphous disk along a number 
of radial string patterns --- this characteristic configuration forms in all 
models and resembles the 
Cat's Cradle. It survives for about 1~Gyr. In the second stage, the disk 
periodically changes its shape from less to more oval and frequently develops
a pair of grand-design spiral arms. The disk shape and the strength of these arms
depends on their relative angle with the major axis of the DM halo. In nearly all
models, the disk grows from inside out and is gas-dominated initially. About
1/3 to 1/2 of the original baryons remains outside the SF region of the disk
or in the halo in form of hot gas at the end of the simulation (see
also Sommer-Larsen 2006).

The star formation (SF) is typically strongly concentrated toward the disk 
midplane. The gaseous component in the disk is puffed up by the feedback from the
stellar evolution. We find that the gas layer narrows substantially when the
SF rate there drops below $\sim 5~{\rm M_\odot~yr^{-1}}$ --- which typically
happens after the first 5~Gyr. The SF delineates various
morphological features --- like early bars, spiral arms and nuclear rings.
Late bars appear gas-poor and do not show much SF. 

We find that parameters which characterize the disk morphology fit the observations
of galactic disks in the nearby universe, e.g., their radial and vertical
scalelengths, shape parameters, bulge-to-disk ratios, etc. The typical radial
scalelength in the disk increases by a factor of 2 over the Hubble time, ending 
within the range of $\sim 2-3$~kpc. During this time the disk grows by more than a 
factor of 10 in mass. 

A caution should be taken when tackling the evolution of a collisionless
component (stars) within the central few 100~pc, specifically the formation of
bulges in disk galaxies. Insufficient timestep resolution can lead to being
unable to resolve the orbits of individual stars --- this leads to energy
non-conservation and overall heating in the region. Such heating has been detected
in the central few 100~pc of our models. It did not affect the large-scale
evolution but could, in principle, modify the nuclear bar strength
as well as increase the dispersion velocities in the disky component of the
bulge. We, therefore, 
postpone the discussion of the bulge formation, the bulge-disk decomposition and the
evolution of characteristic scalelengths in our models to a separate publication
where high-resolution timestep models are presented. 

Essentially, the disks obtained in our simulations range from being bulge-dominated 
to nearly bulgeless. Models with
larger feedback from the stellar evolution form {\it smaller} bulges and vice
versa. Models with shorter timescale for SF formed `fatter' disks but not necessary
more massive bulges. Moreover, reducing the critical density for SF has led to
a distributed SF activity over the inner 15~kpc, as opposed to the inside out disk 
growth in other models, and to `fat' short (small radial scalelength) disks. 

All models are characterized by an extensive bar-forming activity within the 
central few kpc. This is especially true for the initial few Gyr of the disk 
formation. The early bar axial ratios vary in tandem with the mutual bar-halo 
orientation. These bars appear to be gas-rich, channel their gas contents towards 
the central kpc and weaken substantially over this time. In some cases their strength 
was observed to revive, even after a prolonged period of time. In other cases, small 
nuclear bars developed in addition to the primary bars. 
Dynamically, we find that these nuclear bars either corotate or tumble differently 
than the large bars. In all the latter cases, we find that this was preceded by 
a gas inflow to the region and that nuclear bars tumble substantially faster than their 
large-scale counterparts. Some of the nuclear bars appear not to have visible stellar 
components --- they quickly collapse to the center. Nuclear rings and spirals appear 
frequently over the Hubble time and the former ones are robust features in the disk.  
 
Some models show large inflow rates of the gas into the central few kpc leading to
the Jeans instability in the gas and the formation of massive clumps. These clumps
experience dynamical friction within the disk and spiral in before being consumed
by the SF process. Such a regime has been investigated by Shlosman \& Noguchi (1993)
who found that the bar growth is suppressed due to the disk being heated by the
clumps (see also Immeli et al. 2004). Here we confirm that indeed the bars have been 
substantially weakened
by these clumps, or did not form in the extreme cases. 

Our main results indicate that it is possible to form disks which
are in agreement with those in the local universe, e.g.,
with observed sizes, scalelengths, shapes, mass distributions,
etc. Furthermore, we obtain rotation curves that are flat and with the disk/halo
contributions as observed (e.g., Bosma, 2004).
The so-called angular momentum `catastrophe' discussed in section~1
is naturally avoided in this approach. The baryons lose only about 25\% of their angular 
momentum because both the original $J$ present and the feedback from the stellar
evolution reduce their accretion rate onto the disk where most of the $J$ is lost.
We find that in all models the total angular momentum of baryons (stars$+$gas) within 
the disk region closely follows that of the DM there. After the initial $\sim 3$~Gyr,
the baryons rapidly lose their $J$ that flows to the DM until  `equipartition.' 

Moreover, a full range of bulge-dominated to nearly bulgeless disks have been obtained 
as a result of the evolution over the Hubble time. The disks formed in our numerical 
simulations do not appear to suffer from being too compact and having small radial 
or large vertical scalelengths. {\it The growth of the disk cannot be modeled
realiably without resolving its radial scalelength} --- an issue which is difficult
to address at present in cosmological simulations and which is resolved in our
approach. The problem of over-cooling --- when the gas radiates away the energy deposited
by feedback from the stellar evolution (OB stellar radiation-driven winds and SN),
is also addressed here (see section~2).      

Bars provide the strongest impetus for an internal evolution in disk galaxies by 
imposing gravitational torques which trigger the mass and angular momentum redistributions
within the central $\sim 10$~kpc. While prolate DM halos appear to be incompatible
with large-scale stellar bars (El-Zant \& Shlosman 2002; Berentzen et al. 2006), the
growth of disks acts to wash out the equatorial prolateness within, at least, the inner halos
(Berentzen \& Shlosman 2006). The present work confirms this trend --- we
observe a decrease in the halo flatness even well outside the disk, and elimination
of its prolateness within the inner ($< 20$~kpc) halo and reduction in the outer
halo. The disk shape and morphology are affected by the halo shape during the first few
Gyr --- the disks appear oval in contradiction of those observed in the local
universe (e.g., Rix \& Zaritsky 1995). The interaction between the ovally-shaped
disks and prolate halos provide for the spiral driving which clearly depends
on the mutual disk-halo orientations. 

In summary, we have explored different aspects of disk galaxy formation in the
framework of the collapse of an isolated density pertubation embedded in the Hubble 
flow. This allows us to focus on specific processes normally unresolved in the
cosmological simulations. On the other hand, we lose the effect of the substructure
on the model. Specifically, we have aimed at understanding the imprint of various
feedback parameters describing the star formation on the basic charateristics of
the disk-halo system. One should consider this effort as an exploratory one, which requires
a full implementation of realitic initial conditions in tandem with a higher
numerical resolution. 

\acknowledgements
We thank our colleagues, too numerous to list here, for interesting discussions on 
the subject of this work. This research has been partially supported by 
NASA/LTSA 5-13063, NASA/ATP NAG5-10823, HST/AR-10284 (IS), and by NSF/AST 02-06251 
(CH, IS) and by ANR BLAN06-2-134356 (EA). C.H and I.S. are gratefull for the support 
from Observatoire de Marseille during their stay there.

\end{document}